\documentclass[a4paper,11pt]{article}
\usepackage{jcappub} 
\usepackage{lineno}
\usepackage{orcidlink}

\arxivnumber{2211.10464} 

\title{\boldmath Determination of the angular momentum of the Kerr black hole from equatorial geodesic motion.}

\author[a, \orcidlink{0000-0002-0081-3031}]{Laura O. Villegas}
\author[b ]{Eduardo Ramirez-Codiz}
\author[b, \orcidlink{0000-0002-3235-4562}]{V\'ictor Jaramillo}
\author[c, \orcidlink{0000-0002-8603-5209}]{Juan Carlos Degollado}
\author[a, \orcidlink{0000-0002-0496-032X}]{Claudia Moreno}
\author[b, \orcidlink{0000-0003-0295-0053}]{Dar\'io N\'u\~nez}
\author[d, \orcidlink{0000-0003-0962-8390}]{Fernando J. Romero-Cruz}

\affiliation[a]{Departamento de F\'isica,
CUCEI, Universidad de Guadalajara.\\
Blvd. Marcelino García Barragán 1421, C.P. 44430, Guadalajara, Jalisco, M\'exico\\}
\affiliation[b]{Instituto de Ciencias Nucleares, Universidad Nacional Aut\'onoma de M\'exico.\\
Circuito Exterior C.U., A.P. 70-543, M\'exico D.F. 04510, M\'exico\\}
\affiliation[c]{Instituto de Ciencias F\'isicas, Universidad Nacional Aut\'onoma de M\'exico.\\
Apartado. Postal 48-3, 62251, Cuernavaca, Morelos, M\'exico\\}
\affiliation[d]{Instituto Tecnológico Superior de Guanajuato.\\
Carretera Guanajuato-Puentecillas km 10.5 C.P. 36262, Guanajuato, Guanajuato, M\'exico\\}

\emailAdd{laura.villegas8344@alumnos.udg.mx}
\emailAdd{eduardo\_codiz@hotmail.com}
\emailAdd{victor.jaramillo@correo.nucleares.unam.mx}
\emailAdd{jcdegollado@ciencias.unam.mx}
\emailAdd{claudia.moreno@academico.udg.mx}
\emailAdd{fernando.rc@guanajuato.tecnm.mx}

\abstract{We present a method to determine the angular momentum of a black hole based on observations of the trajectories of the bodies in the Kerr spacetime. We use the Hamilton equations to describe the dynamics of a particle and present results for equatorial trajectories, obtaining an algebraic equation for the magnitude of the black hole's angular momentum with coefficients given by observable quantities. We tailor a numerical code to solve the dynamical equations and use it to generate synthetic data. We apply the method in some representative examples, obtaining the parameters of the trajectories as well as the black hole's angular momentum in good agreement with the input data.}

\begin{document}
\maketitle
\flushbottom

\section{Introduction}
\label{sec:introduction}
The precise description of the movement of bodies in the presence of a gravitational field has been the main source for understanding the nature of such interaction. In fact, from the amazing works of Urbain Le Verrier \cite{LeVerrier1846}, theoretically predicting the presence of a new planet, Neptune; to the prediction of precession in the apse of quasi-elliptical trajectories and the deflection of light by the curvature of spacetime \cite{Einstein:1911vc}, a central phenomenon of Einstein's theory of gravitation \cite{Einstein:1914bx}, it has been possible to generate a solid advance in the understanding of the theory of Gravity \cite{Einstein15b}.
Following this line of study, the astronomy group at the University of California (UCLA) has been making precise observations of the motion of the stars in the center of our Galaxy for more than two decades \cite{Gravity_2021b, Ghez:2008ms}. In addition, the Event Horizon Telescope collaboration has accomplished the formidable task of determining the shadow of the object residing in that region \cite{EventHorizonTelescope:2022xnr}.

Taking into account that the solution to Einstein's equations describing a rotating black hole, first derived by Roy Kerr \cite{Kerr:1963ud} is the most plausible and accurate description of the spacetime at the center of our \textit{Via Lactea}, in the present work we propose a method in which, by observing bodies that describe equatorial trajectories in Kerr spacetime, the magnitude of the rotational parameter of the black hole can be inferred. The inference of the such parameter based on observations has been explored in several contexts 
\cite{Guha2010, Hackmann_2008, Kraniotis_2004}, although it has received much less attention than the determination of the other black hole parameter; its mass.
The other parameter in the Kerr spacetime, the mass of the black hole, has received much more attention since the second half of the 20th century, although the first speculations about the existence of black holes were made in 1793 by John Michell \cite{michell}, who used Newton's laws to theorize the characteristics of a star which could retain the light emits. It is possible to measure the mass of Supermassive Black Holes (SMBH) by considering the dynamics of the matter around it \cite{Saglia_2016, Boizelle_2019}. 
The common observational 
technique consists on the measurement of the size of the Broad Line Region (BLR) with the Mapping of the Reverberation (RM) in Active Galactic Nuclei (AGN) \cite{Gravity_2021, Gravity_2022}.
An overview of historical studies of mass and angular moment was included in Appendix \ref{sec:Estimation}.

In this paper, we consider that the mass of the black hole under study has been determined. In order to test our proposal, we used the Hamiltonian formulation and develop a numerical code to evolve the geodesic motion \cite{walecka2007introduction} using synthetic data such as mass, distance and position of the apsides.
These actions allow us to generate trajectories in the Kerr spacetime and produce data which is used to prove the viability of the method presented. This problem is interesting because so far there is no real observational measurement of angular momentum and this work could help astrophysicists trying to understand it.
The manuscript is organized as follows:  
In section \ref{metric} we  describe the Kerr black hole in
Kerr-Schild coordinates, and present the expressions for the components of the four velocity which play a central role in our procedure. The derivation of the equations of motion is presented in the Appendix \ref{Sec:Eqs-motion}. In the Appendix \ref{sec:Carter} we present an alternative derivation of the Carter's constant, a conserved quantity in the Kerr space. Through out the work we express all the functions in terms of characteristic quantities for the distance, time, and mass, which allows to give a dimensionless description and directly recover physical data when needed. In section \ref{sec:procedure}, we present the details of the method to determine the angular momentum from the velocity components expressing the system of equations for such components with coefficients described in terms of known quantities, finally arriving to a sixth degree algebraic equation for the angular parameter of the black hole. We also present the case of a Schwarzschild background and obtain a new expression for determining the mass of such black hole, in Appendix \ref{sec:BH_Mass} we use the code to evaluate the precision in the determination of the black hole mass using this expression. Next, in section, \ref{sec:mock_data}, we use the code to generate synthetic data. We present cases for two super massive black holes, with examples for three different values of the angular momentum, the quantity to be determined, and using trajectories with apsides near and far to the black hole horizon. 
In section \ref{sec:Non_equatorial} we give a brief discussion on how it could be possible to extend the algorithm to non-equatorial trajectories and in the final section \ref{sec:final_comms}, we present the conclusions and suggestions for future work. Finally, in Appendix \ref{sec:Mercury} we use data from our Solar System as a test.

\section{Kerr metric}
\label{metric}
%
The line element of the Kerr spacetime in Kerr-Schild coordinates \cite{Kerr:1963ud, ryder_2009} is  
\begin{eqnarray}
 d\hat{s}^2 &=&  -\left(1 - \frac{2\sigma^2}{\hat{r} \, \Delta_{0+}}\right) \, c^2\,\frac{d\,\hat{t}^2}{\sigma^2} + \frac{4\,\sigma}{\hat{r}\Delta_{0+}}\,c\,d\,\hat{t}\,d\,\hat{r} - \frac{4\,\sigma^3\,{j}\,\sin^2\theta}{\hat{r}\,\Delta_{0+}}\,d\hat{t}\,d\phi + 
 \left(1 +\frac{2\,\sigma^2}{\hat{r}\Delta_{0+}}\right)\,d\hat{r}^2  
 \\
&-& 2\,\sigma^2 {j}\left(1+\frac{2\,\sigma^2}{\hat{r}\,\Delta_{0+}}\right)\,\sin^2\theta\,
d\hat{r}\,d\phi + \hat{r}^2 \Delta_{0+}\,d\theta^2 + \hat{r}^2\left(1 + \frac{\sigma^4\, {j}^2}{\hat{r}^2} + \frac{2\, {j}^2\, \sigma^6\,\sin^2\theta}{\hat{r}^3 \Delta_{0+}}\right)\,\sin^2\theta\,d\phi^2, \nonumber \label{eq:ds2K}
\end{eqnarray}
where 
\begin{eqnarray}
\sigma^2&=&\frac{G\,M}{c^2\,R_0}, \label{def:sigma}\\
a&=&\frac{J\,c}{M}=\frac{J_{\rm max}\,j\,c}{M}=\sigma^2\,R_0\,j, \\
J_{\rm max}&=&\frac{M^2\,G}{c}, \\
\Delta_{0\pm} &=& 1 \pm \frac{\sigma^4\,{j}^2 \cos^2\theta}{\hat{r}^2},\label{Delta0}
\end{eqnarray}
with $J_{\rm max}$ the maximal angular momentum of the black hole, {\it i.e.} the angular momentum of the extreme black hole; $R_0$ is a characteristic distance of the system, so that the radial coordinate can be written in terms of such distance and a dimensionless distance, $\hat{r}$: $r=R_0\,\hat{r}$. We also defined a fiducial time, $T_0=\frac{R_0}{c}$, so that we express the proper and coordinate time as $\tau=T_0\,\hat{\tau}$ and $t=T_0\,\hat{t}$ respectively.

\subsection{Spatial components of the four velocity and trajectories}

As explained in the Appendix \ref{Sec:Eqs-motion} once the equations of motion for the Hamiltonian variables are obtained, it is straightforward to derive the expressions for the temporal and spatial components of the four velocity. These are:
\begin{eqnarray}
\Delta\,{\hat{u}}^0 &=& \frac{2\,\sigma^2}{\hat{r}}\,\hat{u}^r - \frac{2\,\sigma^4\,j}{{\Delta_{0+}\,\hat{r}}^3}\,\hat{L}_\phi \label{eq:u0}\\
&& - \frac{\sqrt{1+2\,\hat{E}_N}}{\Delta_{0+}} \left[1 + \frac{\sigma^4\,j^2}{{\hat{r}}^2}\,\left(1 + \frac{2\,\sigma^2}{{\hat{r}}} + \Delta\,\cos^2\theta\right)\right] \,, \nonumber
\end{eqnarray}
\begin{eqnarray}
\frac{{{\Delta_{0+}}^2}}{2}\,({\hat{u}^r})^2&=&\hat{E}_N + \frac{\sigma^2}{\hat{r}} - \frac{{\hat{L}_\phi}^2 - 2\,\sigma^4\,j^2\,\hat{E}_N}{2\,{\hat{r}}^2}  \\
&& + \frac{\sigma^2\,\left(\hat{L}_\phi - \sigma^2\,j\,\sqrt{1+2\,\hat{E}_N}\right)^2}{{\hat{r}}^3} - \frac{\Delta\,\hat{C}^2}{2\,{\hat{r}}^2},  \label{eq:ur} \nonumber\\
{{\Delta_{0+}}^2}\,({\hat{u}^\theta})^2&=&\frac{\hat{C}^2 - \cot^2\theta\,{\hat{L}_\phi}^2 + 2\,j^2\sigma^4\,\hat{E}_N\,\cos^2\theta}{\hat{r}^4}, \label{eq:uth} 
\end{eqnarray}

\begin{eqnarray}
\hat{r}^2\,\Delta_{0+}\,\hat{u}^\phi&=&\frac{\hat{L}_\phi}{\sin^2\theta} + \frac{j\,\sigma^2 \,\Delta_{0+}}{\Delta}\,\hat{u}^r \\
&& + \frac{j\,\sigma^2}{\Delta} \left( \frac{2\,\sigma^2}{\hat{r}}\,\sqrt{1+2\,\hat{E}_N}-\frac{j\,\sigma^2}{\hat{r}^2}\,\hat{L}_\phi\right), \label{eq:uf}\nonumber
\end{eqnarray}
 where $\Delta=1-\frac{2\,\sigma^2}{\hat{r}} + \frac{\sigma^4\,j^2}{{\hat{r}}^2}$, $\hat{L}_\phi$ is the conserved azimuthal angular momentum, $\hat{C}$ is the Carter's constant \cite{Carter1968HamiltonJacobiAS}, and $\hat{E}_N$ is a conserved quantity, related to the conserved quantity associated with the temporal symmetry of the spacetime $\hat{p}_0$, see Eq.~(\ref{eq:p0-EN}) which can be identified with the total energy when the Newtonian limit is taken. In these expressions, we have used the following definitions $\Delta\,\hat{p}_r=\Delta_{0+}\,\hat{u}^r - \frac{\sigma^2\,j}{{\hat{r}}^2}\,\hat{L}_\phi + \frac{2\,\sigma^2}{{\hat{r}}}\,\sqrt{1+2\,\hat{E}_N}$.
 
The Carter constant, plays  a similar  role in the Kerr spacetime as the magnitude of the total angular momentum, plays in the Schwarzschild motion. Indeed, for $j=0$, the total angular momentum, $\hat{L}_T$, is related to the Carter constant as ${\hat{L}_T}^2=\hat{C}^2 + {\hat{L}_\phi}^2$. The conserved quantities are already dimensionless by means of  $E_N=m\,c^2\,\hat{E}_N$ and $C=m\,c\,R_0\,\hat{C}, \,L_\phi=m\,c\,R_0\,\hat{L}_\phi$.

The fact that the total angular momentum is not a conserved quantity in the Kerr spacetime, implies that the trajectory does not remain in a plane, which is a distinguishing feature of the Kerr metric. 

\begin{figure*}
    \centering
    \includegraphics[scale=0.5]{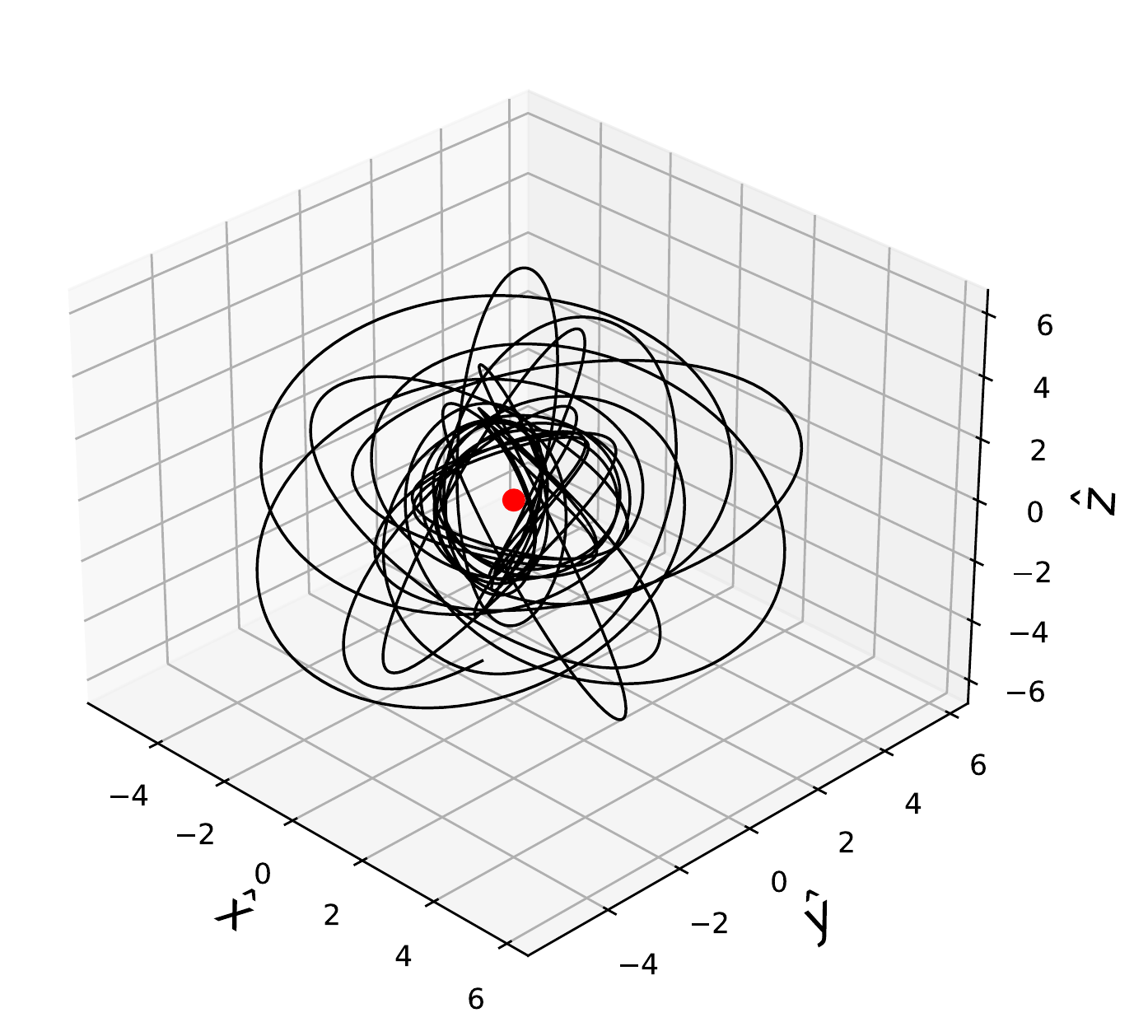} \includegraphics[scale=0.5]{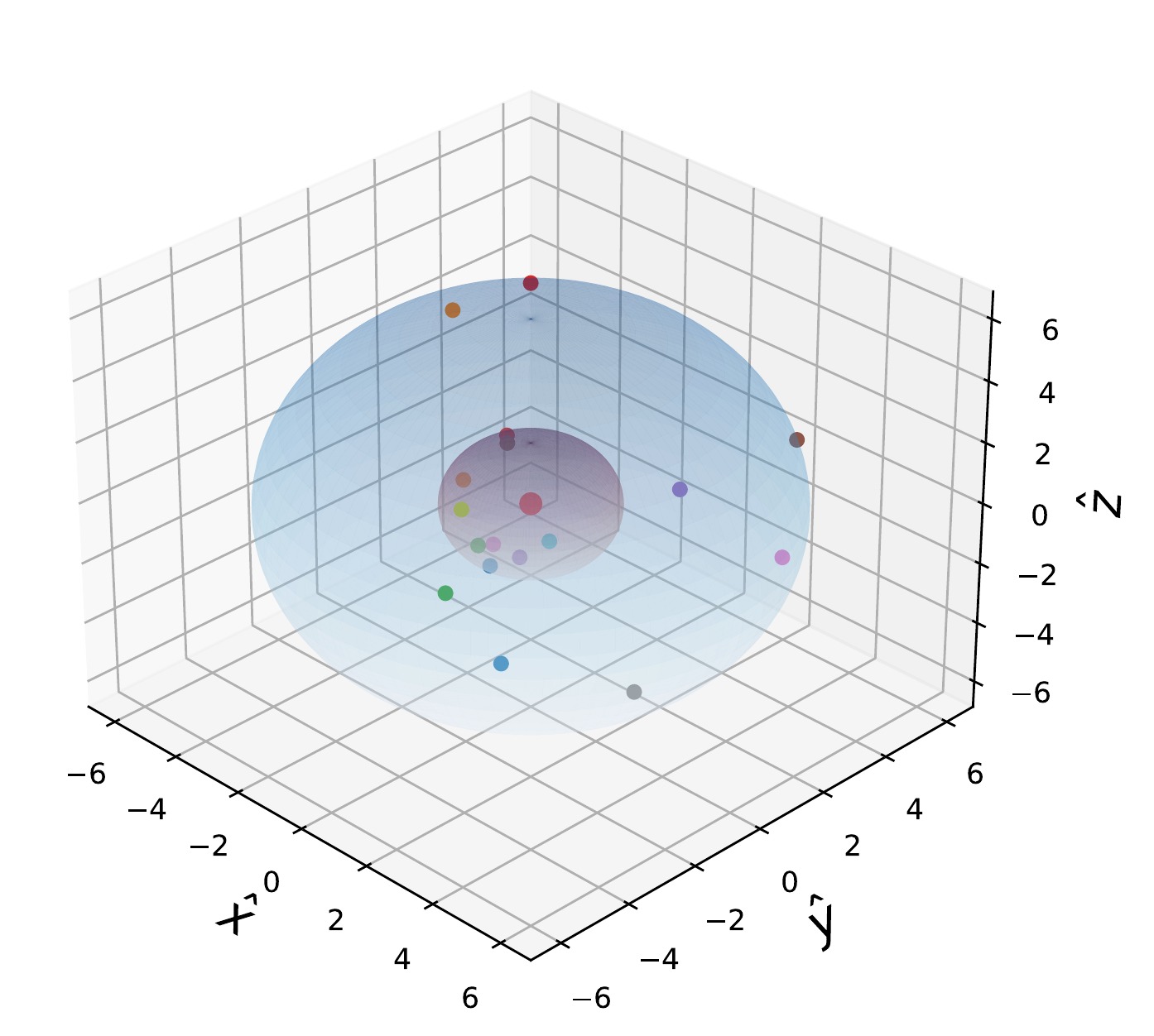}
    \caption{Left plot shows a trajectory in a Kerr spacetime with $j=0.98$, a mass of $M=5.07\,\times\,10^7\,M_\odot$, and $R_0=1\,{\rm AU}$, showing that trajectories do not remain on a plane.  In the plot on the right, we present the position where the test particle reaches the apsides. The concentric spheres correspond to the values of the radial apsides, located at $\hat{r}_1=2, \, \hat{r}_2= 5.5$ . The position of the apsides changes with each turn. The axis are described in terms of the dimensionless Cartesian coordinates $\hat{x}, \hat{y}, \hat{z}$.}
    \label{fig:traj_j0p98}
\end{figure*}

In Fig.~\ref{fig:traj_j0p98}, it is shown an example of a typical orbit in Kerr; notice how the trajectories do not remain on a plane, and we also show the position of the apsis which, while occurring always at the fixed radial values, $\hat{r}_1, \hat{r}_2$ their corresponding angular positions change due to the fact that the precession occurs in of both angles $\theta$ and $\phi$ {\it i.e.} the Lense-Thirring precession.
Notice that such double precession is presented in the bounded trajectory of any test object freely moving in the Kerr spacetime; their magnitude decreases as the trajectory of the object is farther away from the black hole.

\section{Black hole angular momentum determination} 
\label{sec:procedure}

To determine the angular momentum of the black hole we assume that the mass of the black hole \cite{Ghez:2008ms} is known, as well as the physical distances of the apses, $\hat r_1, \hat r_2$. With these data obtained by observations, it is possible to fix the distance $R_0$ and therefore the value of $\sigma$, Eq.~(\ref{def:sigma}). Furthermore, we consider the angle of inclination of the observed orbit with respect to the direction of the angular momentum vector of the black hole \cite{EventHorizonTelescope:2022urf, EventHorizonTelescope:2022xqj}. The values of the physical parameters that we took were implemented in two examples: first we will consider an object with the mass similar to that of Sagittarius A*. The second example will be an object with the mass such that the event horizon radius is 1 AU. In both cases we will take two particle trajectories around these objects, one close and one far, to observe the relativistic behavior and its importance in the precession of the orbit.

With these assumptions, we use the fact that the radial component of the velocity, Eq.~(\ref{eq:ur}), evaluated at each one of the apsis is equal to zero. This gives us two equations for four unknowns, namely $\hat{E}_N, \, \hat{L}_\phi, \, \hat{C}^2$ and $j$. Two more equations are needed.
Noticing that at the apsides, as the radial component of the velocity is zero, the spatial displacement is determined only by the angular components of the velocity. With this fact in mind, the determination of the spin is based in the assumption that the angular components of the velocity can be determined by means of observations of the angular displacements near an apsis.

Such idea can be directly applied for the case of equatorial trajectories, where the angular displacement can be directly associated with the azimuthal angle, and we can use the definition of the angular velocity, $\hat{\Omega}$:
\begin{equation}
\hat{\Omega}=\frac{d\,\phi}{d\,\hat{t}}=\frac{\hat{u}^\phi}{\hat{u}^0}, \label{eq:Omega0}    
\end{equation}
and using the expressions for the azimuthal and the temporal component of the four velocity, Eqs.~(\ref{eq:ham_t}, \ref{eq:ham_f}), evaluated at the apsis (where $\hat{u}^r=0$) we obtain that
\begin{equation}
\hat{\Omega}|_{ap}=\frac{\Delta_S(\hat{r}_a)\,\hat{L}_\phi + \frac{2\,\sigma^4\,j}{\hat{r}_a}\,\sqrt{1+2\,\hat{E}_N}}{{\hat{r}_a}^2\,\left(\Delta_3(\hat{r}_a)\,\sqrt{1+2\,\hat{E}_N} - \frac{2\,\sigma^4\,j}{{\hat{r}_a}^3}\,\hat{L}_\phi\right)},    
\end{equation}
with $\Delta_3(\hat{r})=1+\frac{\sigma^4\,j^2}{{\hat{r}}^2\,\left(1+\frac{2\,\sigma^2}{\hat{r}}\right)}, \, \Delta_S(\hat{r})=1-\frac{2\,\sigma^2}{\hat{r}}$, and $\hat{r}_a$ stands for the radius of an apsis. We consider that $\hat{\Omega}|_{ap}$ can be determined by observations. From this last equation, we can express the azimuthal angular momentum as:
\begin{equation}
\hat{L}_\phi={\hat{r}_a}^2\,\sqrt{1+2\,\hat{E}_N}\,
\frac{\hat{\Omega}|_{ap}\,\Delta_3(\hat{r}_a)- \frac{2\,\sigma^4\,j}{{\hat{r}_a}^3}}{\Delta_S(\hat{r}_a) - \frac{2\,\sigma^4\,j}{{\hat{r}_a}}\,\hat{\Omega}|_{ap}}.  \label{eq:LfK_ecu}  \end{equation}
These expressions allow us to conclude that the angular velocity is the same at the apheaster, $\hat{r}_1$, and also has the same value for the periaster, $\hat{r}_2$. Moreover, these two values are related to each other as:
\small{
\begin{equation}
\hat{\Omega}|_{\hat{r}_1}=\frac{F_1\,\hat{\Omega}|_{\hat{r}_2} + 2\,\sigma^4\,j\,\hat{r}_2\,\left(1 - \frac{\hat{r}_1}{\hat{r}_2}\right)}{F_2 + 2\,\sigma^4\,j\,{\hat{r}_2}^3\,\left(1 - \frac{\hat{r}_1}{\hat{r}_2}\right)\,\left(1 + \frac{\hat{r}_1}{\hat{r}_2} + \left(1 - \frac{\hat{r}_1}{\hat{r}_2}\right)^2 + \frac{\sigma^4\,j^2}{{\hat{r
_2}^2}}\right)\,\hat{\Omega}|_{\hat{r}_2}},  \label{eq:OmegasK_ecu}  
\end{equation}}
where 
\begin{eqnarray}
 F_1&=&{\hat{r}_1}\,{\hat{r}_2}^3\,\left(1-\frac{2\,\sigma^2}{\hat{r_1}} + \frac{\sigma^4\,j^2}{{\hat{r}_2}^2}\,\left(1 - 2\,\sigma^2\,\left(\frac{1}{{\hat{r}_1}} - \frac{1}{{\hat{r}_2}}\right)\right)
\right) \,, \\
F_2&=&{\hat{r}_1}^3\,{\hat{r}_2}\,\left(1-\frac{2\,\sigma^2}{\hat{r_2}} + \frac{\sigma^4\,j^2}{{\hat{r}_1}^2}\,\left(1 + 2\,\sigma^2\,\left(\frac{1}{{\hat{r}_1}} - \frac{1}{{\hat{r}_2}}\right)\right)
\right) \,.   
\end{eqnarray}

We can use expression in Eq.~(\ref{eq:LfK_ecu}) in the equations for the radial component of the velocity that, evaluated at $\hat{r}_1$ and $\hat{r}_2$, in the equatorial case that we are considering, takes the form:
\begin{eqnarray}
\hat{E}_N + \frac{\sigma^2}{\hat{r}_1} - \frac{{\hat{L}_\phi}^2 - 2\,\sigma^4\,j^2\,\hat{E}_N}{2\,{\hat{r}_1}^2} + 
\frac{\sigma^2\,\left(\hat{L}_\phi - \sigma^2\,j\,\sqrt{1+2\,\hat{E}_N}\right)^2}{{\hat{r}_1}^3}=0,  \label{eq:ur2K_r1_ecu}\\
\hat{E}_N + \frac{\sigma^2}{\hat{r}_2} - \frac{{\hat{L}_\phi}^2 - 2\,\sigma^4\,j^2\,\hat{E}_N}{2\,{\hat{r}_2}^2} + 
\frac{\sigma^2\,\left(\hat{L}_\phi - \sigma^2\,j\,\sqrt{1+2\,\hat{E}_N}\right)^2}{{\hat{r}_2}^3}=0\,.  \label{eq:ur2K_r2_ecu}
\end{eqnarray}

Let us consider that we made observations of the angular and temporal displacements at $\hat{r}_2$, and used it in these last equations, 
\begin{eqnarray}
&&\hat{E}_N + \frac{\sigma^2}{\hat{r}_1} - \frac{\left({\hat{r}_2}^2\,\sqrt{1+2\,\hat{E}_N}\,
\frac{\hat{\Omega}|_{\hat{r}_2}\,\Delta_3(\hat{r}_2)- \frac{2\,\sigma^4\,j}{{\hat{r}_2}^3}}{\Delta_S(\hat{r}_2) - \frac{2\,\sigma^4\,j}{{\hat{r}_2}}\,\hat{\Omega}|_{\hat{r}_2}}\right)^2 - 2\,\sigma^4\,j^2\,\hat{E}_N}{2\,{\hat{r}_1}^2} \nonumber \\
&& + \frac{\sigma^2\,\left(1+2\,\hat{E}_N\right)}{{\hat{r}_1}^3}\,\left({\hat{r}_2}^2\,
\frac{\hat{\Omega}|_{\hat{r}_2}\,\Delta_3(\hat{r}_2)- \frac{2\,\sigma^4\,j}{{\hat{r}_2}^3}}{\Delta_S(\hat{r}_2) - \frac{2\,\sigma^4\,j}{{\hat{r}_2}}\,\hat{\Omega}|_{\hat{r}_2}} - \sigma^2\,j\right)^2=0,  \nonumber \\
\\
&&\hat{E}_N + \frac{\sigma^2}{\hat{r}_2} - \frac{\left({\hat{r}_2}^2\,\sqrt{1+2\,\hat{E}_N}\,
\frac{\hat{\Omega}|_{\hat{r}_2}\,\Delta_3(\hat{r}_2)- \frac{2\,\sigma^4\,j}{{\hat{r}_2}^3}}{\Delta_S(\hat{r}_2) - \frac{2\,\sigma^4\,j}{{\hat{r}_2}}\,\hat{\Omega}|_{\hat{r}_2}}\right)^2 - 2\,\sigma^4\,j^2\,\hat{E}_N}{2\,{\hat{r}_2}^2} \nonumber\\
&& + \frac{\sigma^2\,\left(1+2\,\hat{E}_N\right)}{{\hat{r}_2}^3}\,\left({\hat{r}_2}^2\,
\frac{\hat{\Omega}|_{\hat{r}_2}\,\Delta_3(\hat{r}_2)- \frac{2\,\sigma^4\,j}{{\hat{r}_2}^3}}{\Delta_S(\hat{r}_2) - \frac{2\,\sigma^4\,j}{{\hat{r}_2}}\,\hat{\Omega}|_{\hat{r}_2}} - \sigma^2\,j\right)^2=0. 
\end{eqnarray}
A straightforward manipulation of the second equation allows us to express $\hat{E}_N$ as:
\begin{equation}
\hat{E}_N=-\frac12\,\frac{\hat{\Omega}|_{\hat{r}_2}\,\left(A - \frac{8\,\sigma^6\,j}{{\hat{r}_2}^4}\right) - \frac{2\,\sigma^2\,\Delta_S(\hat{r}_2)}{{\hat{r}_2}^3}}{\hat{\Omega}|_{\hat{r}_2}\,\left(B - \frac{4\,\sigma^4\,j}{{\hat{r}_2}^3}\right) - \frac{2\,\sigma^2\,\Delta_S(\hat{r}_2)}{{\hat{r}_2}^2}}, \label{eq:ENK_ecu}    
\end{equation}

where we have defined 

\begin{eqnarray}
 A&=&\hat{\Omega}|_{\hat{r}_2}\,\left(1 + \frac{\sigma^4\,j^2}{{\hat{r}_2}^2}\,\left(1 + \frac{2\,\sigma^2}{\hat{r}_2}\,\left(1 + \frac{2\,\sigma^2}{\hat{r}_2}\right)\right)\right) \,, \\ 
 B&=&\hat{\Omega}|_{\hat{r}_2}\,\left(1 + \frac{\sigma^4\,j^2}{{\hat{r}_2}^2}\,\left(1 + \frac{2\,\sigma^2}{\hat{r}_2}\right)\right) \,.
\end{eqnarray}

Using this expression for $\hat{E}_N$ in the radial equation evaluated at $\hat{r}_1$, we finally obtain an equation only for the angular momentum $j$. It is an algebraic equation of sixth degree in $j$ (is identically zero for the circular case, as we have left indicated): 

\begin{equation}
\left(\hat{r}_1-\hat{r}_2\right)\,\left(a_6\,j^6 + a_5\,j^5 + a^4\,j^4 + a_3\,j^3 + a_2\,j^2 + a^1\,j + a_0\right)=0, \label{eq:alg_j}    
\end{equation}

with the next definitions:
{\small
\begin{eqnarray}
a_6&=&8\,{\hat{r}_2}^2\,\left({\hat{\Omega}|_{\hat{r}_2}}\right)^4\,\sigma^{18}, \\
a_5&=&8\,{\hat{r}_2}^4\,\left({\hat{\Omega}|_{\hat{r}_2}}\right)^3\,\sigma^{14}\,\left(1-\frac{4\,\sigma^2}{\hat{r}_2}\right), \nonumber \\
a_4&=&-2\,{\hat{r}_2}^2\,\left({\hat{\Omega}|_{\hat{r}_2}}\right)^2\,\sigma^{10}\,\left(2\,\sigma^2\,{\hat{r}_1}\,{\hat{r}_2}^3\,\left(1 - \frac{4\,\sigma^2}{{\hat{r}_1}} + \frac{\hat{r}_1}{\hat{r}_2} \right. \right. \nonumber \\
&&\left. \left. + \frac{2\,\sigma^2\,\hat{r}_1}{{\hat{r}_2}^2} + \frac{4\,\sigma^4\,\hat{r}_1}{{\hat{r}_2}^3} \right)\,\left({\hat{\Omega}|_{\hat{r}_2}}\right)^2 - {\hat{r}_2}^3\,\left(1-\frac{12\,\sigma^2}{{\hat{r}_2}} + \frac{24\,\sigma^4}{{\hat{r}_2}^2}\right)\right), \nonumber \\
a_3&=&-4\,{\hat{r}_2}^2\,{\hat{\Omega}|_{\hat{r}_2}}\,\sigma^{8}\,\left(\hat{r}_1\,{\hat{r}_2}^4\,\left(\Delta_2\left({\hat{r}_2}\right) - \frac{4\,\sigma^2}{{\hat{r}_1}} + \frac{12\,\sigma^4}{{\hat{r}_1}\,{\hat{r}_2}} + \frac{\hat{r}_1}{\hat{r}_2} \right.\right. \nonumber\\
&&\left. \left. - \frac{16\,\hat{r}_1\,\sigma^6}{{\hat{r}_2}^4} \right)\,\left({\hat{\Omega}|_{\hat{r}_2}}\right)^2 + {\hat{r}_2}^3\,\Delta_S\left(\hat{r}_2\right)\,\left(1 - \frac{4\,\sigma^2}{{\hat{r}_2}}\right)\right), \nonumber \\
a_2&=&{\hat{r}_2}^2\,\sigma^4\,\left(4\,{\hat{r}_2}^5\,\hat{r}_1\,\,\sigma^4\,\left(1+\frac{\hat{r}_1}{\hat{r}_2}\right)\,\Delta_S\left(\hat{r}_1\right)\left({\hat{\Omega}|_{\hat{r}_2}}\right)^4 \right. \nonumber \\ 
&&\left. + \,\Delta\left(\hat{r}_2\right)\hat{r}_1\,{\hat{r}_2}^4\left(\Delta\left(\hat{r}_2\right) - \frac{4\sigma^2}{{\hat{r}_1}} + \frac{24\sigma^4}{{\hat{r}_1}{\hat{r}_2}} + \frac{\hat{r}_1}{\hat{r}_2} - \frac{48\hat{r}_1\sigma^6}{{\hat{r}_2}^4}\right)\left({\hat{\Omega}|_{\hat{r}_2}}\right)^2 \right. \nonumber\\
&& \left.- 2\,{\hat{r}_2}^3\sigma^2\,\Delta\left(\hat{r}_2\right)\right), \nonumber \\
a_1&=&-4\,{\hat{r}_2}^5\,\hat{\Omega}|_{\hat{r}_2}\sigma^4\,\Delta_S\left(\hat{r}_2\right)\,\left({\hat{r}_2}^5\,\hat{r}_1\,\Delta_S\left(\hat{r}_1\right)\,\left(1+\frac{\hat{r}_1}{\hat{r}_2}\right)\,\left({\hat{\Omega}|_{\hat{r}_2}}\right)^2 \right. \nonumber \\ 
&&\left. + \,{\hat{r}_2}^4\,\,\Delta_S\left(\hat{r}_2\right)\,\left(1-\frac{4\,\sigma^2\,{\hat{r}_1}^2}{\hat{r}_2}\right)\right), \nonumber \\
a_0&=&-{\hat{r}_2}^4\,\left({\Delta\left(\hat{r}_2\right)}\right)^2\,\left({\hat{r}_2}^4\,{\hat{r}_1}\,\Delta\left(\hat{r}_2\right)\,\left(\hat{r}_1+\hat{r}_2\right)\,\left({\hat{\Omega}|_{\hat{r}_2}}\right)^2 \right. \nonumber\\
&&\left. - 2{\hat{r}_1}^2\,\hat{r}_2\,\sigma^2\,\left(\Delta\left(\hat{r}_2\right)\right)^2\right)\,. \nonumber
\end{eqnarray}
}

In this way, given $\hat{r}_1, \, \hat{r}_2, \, \sigma$, and $\hat{\Omega}|_{\hat{r}_2}$, the algebraic equation for $j$ can be solved, thus determining the angular momentum of the black hole. Finally, one determines the energy $\hat{E}_N$ and the azimuthal angular momentum $\hat{L}_\phi$ which determines the trajectory.

Before proceeding to the discussion of the method in concrete examples, we want to present the Schwarzschid case under these ideas. In principle, as we will show, the algorithm determines the value of $j$ including the case for $j=0$, but let us consider the case that we know that we deal with a Schwarzschild black hole and that we can observe the trajectory of a body in its vicinity and measure the distances of the apsides, $\hat{r}_1, \hat{r}_2$, thus determine the distance, $R_0$ and that we can  determine $\hat{\Omega}|_{ap}$. We orient our reference system such that the equatorial plane coincides with the orbit, so that we can use Eq.~(\ref{eq:LfK_ecu}) with $j=0$ to express the azimuthal momentum in terms of the angular velocity:
\begin{equation}
\hat{L}_\phi={\hat{r}_a}^2\,\sqrt{1+2\,\hat{E}_N}\,
\frac{\hat{\Omega}|_{ap}}{1-\frac{2\,\sigma^2}{\hat{r}_a}},  \label{eq:LfS}  
\end{equation}
from which we obtain the following relation between the angular velocities measured at each apsis:
\begin{equation}
\hat{\Omega}|_{\hat{r}_1}=\left(\frac{\hat{r}_2}{\hat{r}_1}\right)^2\frac{1-\frac{2\,\sigma^2}{\hat{r}_1}}{1-\frac{2\,\sigma^2}{\hat{r}_2}}\,\hat{\Omega}|_{\hat{r}_2}.  \label{eq:OmegasS}  
\end{equation}
Using this information in the equations for the radial component of the velocity evaluated at the apsis, Eqs.~(\ref{eq:ur2K_r1_ecu}, \ref{eq:ur2K_r2_ecu}) with $j=0$, we obtain the following system of equations for the energy $\hat{E}_N$ and $\sigma^2$, that is, the mass:
\begin{eqnarray}
2\,\hat{E}_N + \frac{2\,\sigma^2}{\hat{r}_1} - \left(1+2\,\hat{E}_N\right)\,\left(\frac{1-\frac{2\,\sigma^2}{\hat{r}_1}}{1-\frac{2\,\sigma^2}{\hat{r}_2}}\right)\,\left(\frac{{\hat{r}_2}^4}{{\hat{r}_1}^2}\right)\,\hat{\Omega}|_{\hat{r}_2}&=&0,  \label{eq:ur2S_r1}\\
2\,\hat{E}_N + \frac{2\,\sigma^2}{\hat{r}_2} - \left(1+2\,\hat{E}_N\right)\,\left(\frac{{\hat{r}_2}^2}{1-\frac{2\,\sigma^2}{\hat{r}_2}}\right)\,\left(\frac{{\hat{r}_2}^4}{{\hat{r}_1}^2}\right)\,\hat{\Omega}|_{\hat{r}_2}&=&0,  \label{eq:ur2S_r2}
\end{eqnarray}
which can be solved for the energy $\hat{E}_N$ and $\sigma$ in terms of the observed quantities:
\small{
\begin{eqnarray}
&&\hat{E}_N=\left[\frac{\pm{\hat{r}_2}^4\,\left(\hat{\Omega}|_{\hat{r}_2}\right)^2}{2\,{\hat{r}_1}^2\,\left(\left(\left(\hat{\Omega}|_{\hat{r}_2}\right)^2\,\left(2\,{\hat{r}_1}^2 + {\hat{r}_2}^2 + {\hat{r}_1}\,{\hat{r}_2}\right) - \left(\frac{\hat{r}_1}{\hat{r}_2}\right)^2 \right) + A_S\right)}\right]\times \nonumber \\
&&\,\left[\left(\frac{\hat{r}_1}{\hat{r}_2}\right)^2\,\left({\hat{r}_2}^2\,\left(\hat{\Omega}|_{\hat{r}_2}\right)^2\,\left(1+\frac{\hat{r}_2}{\hat{r}_1}\right)^2 + 1 - \frac{\hat{r}_1}{\hat{r}_2}\right) + \left(1+\frac{\hat{r}_1}{\hat{r}_2}\right)\,A_S\right], \label{eq:EN_Schw} \\
&&\sigma^2=\frac{\hat{r}_2}{4}\,\left(1 + \left(\frac{\hat{r}_2}{\hat{r}_1}\right)\,{\hat{r}_2}^2\,\left(\hat{\Omega}|_{\hat{r}_2}\right)^2\,\left(1+\frac{\hat{r}_2}{\hat{r}_1}\right) \mp  \left(\frac{\hat{r}_2}{\hat{r}_1}\right)^2\,A_S\right), \label{eq:sigma_Schw}
\end{eqnarray}}
with 
\small{
\begin{equation}
A_S=\sqrt{\left(\hat{\Omega}|_{\hat{r}_2}\right)^2\,\left(1 + \frac{\hat{r}_1}{\hat{r}_2}\right)\,\left({\hat{r}_2}^4\,\left(\hat{\Omega}|_{\hat{r}_2}\right)^2\,\left(1+\frac{\hat{r}_1}{\hat{r}_2}\right) + 2\,{\hat{r}_1}^2\,\left(1-2\,\frac{\hat{r}_1}{\hat{r}_2}\right) \right) + \left(\frac{\hat{r}_1}{\hat{r}_2}\right)^4}.    
\end{equation}}

In this way, considering that we fix the fiducial distance of the system, $R_0$, knowing $\sigma^2$, with Eq.~(\ref{def:sigma}) we can determine the mass of the black hole in the Schwarzschild case from quasi-elliptical trajectories. 

In the case of circular orbits, $\hat{r}_1=\hat{r}_2=\hat{R}$, our expressions takes the form $\hat{E}_N=-\frac12, \, \sigma^2=\frac{\hat{R}}{2}$, for the upper sign (only valid when the radius coincides with the horizon) and, for the lower one:
\begin{eqnarray}
\hat{E}_N&=&-\hat{R}^2\,\left(\hat{\Omega}|_{\hat{R}}\right)^2\,\frac{1 - 4\,\hat{R}^2\,\left(\hat{\Omega}|_{\hat{R}}\right)^2}{2\,\left(1 - 3\,\hat{R}^2\,\left(\hat{\Omega}|_{\hat{R}}\right)^2\right)}, \label{eq:EN_Schw_c}\\
\sigma^2&=&\hat{R}^3\,\left(\hat{\Omega}|_{\hat{R}}\right)^2, \label{eq:sigma2_Schw_c}
\end{eqnarray}
which are the usual expressions for the energy and the mass in this case. In practice, however, the expressions with the upper sign are the ones that give the correct value of the mass and energy in concrete examples.

In this way, our procedure in the Schwarzschild case implies that knowing the apsis distances, and measuring the angular velocity at any apsis, we can determine the magnitude of the total angular momentum, $\hat{L}_T$, the energy of the object and the mass of the black hole. For completeness, we mention that the Keplerian case within this approach allows to determine the energy of the moving body and the mass as:
\begin{eqnarray}
\hat{E}_N&=&-\frac{{\hat{\Omega}|_{\hat{r}_2}}^2\,{\hat{r}_2}^3}{2\,\hat{r}_1}, \\
\sigma^2&=&\frac{{\hat{\Omega}|_{\hat{r}_2}}^2\,{\hat{r}_2}^3\,\left(\hat{r}_2+\hat{r}_1\right)}{2\,\hat{r}_1}. \label{eq:sigma_K}
\end{eqnarray}

\section{Synthetic data}
\label{sec:mock_data}

We developed a fourth order Runge Kutta code to solve the dynamical equations Eqs.~(\ref{eq:ham_t}-\ref{eq:ham_Lth}). We have been able to reproduce the observed precession of Mercury in the Solar System \cite{Park2017} for several values of the parameter$j$ the results are presented in the Appendix \ref{sec:Mercury}. We know that the advance of Mercury's perihelion is very small, just 42.98 ${\rm arcsec}$ per century \cite{Magnan2007}; our code calculated this advance with a precision of $10^{-7}$. The accuracy of our results gives confidence that the code can accurately describe the dynamics of the bodies in the Kerr spacetime, and include the effects of the Schwarzschild spacetime or in a Keplerian description, as limiting cases.

\begin{figure}[ht]
    \centering
    \includegraphics[scale=0.5]{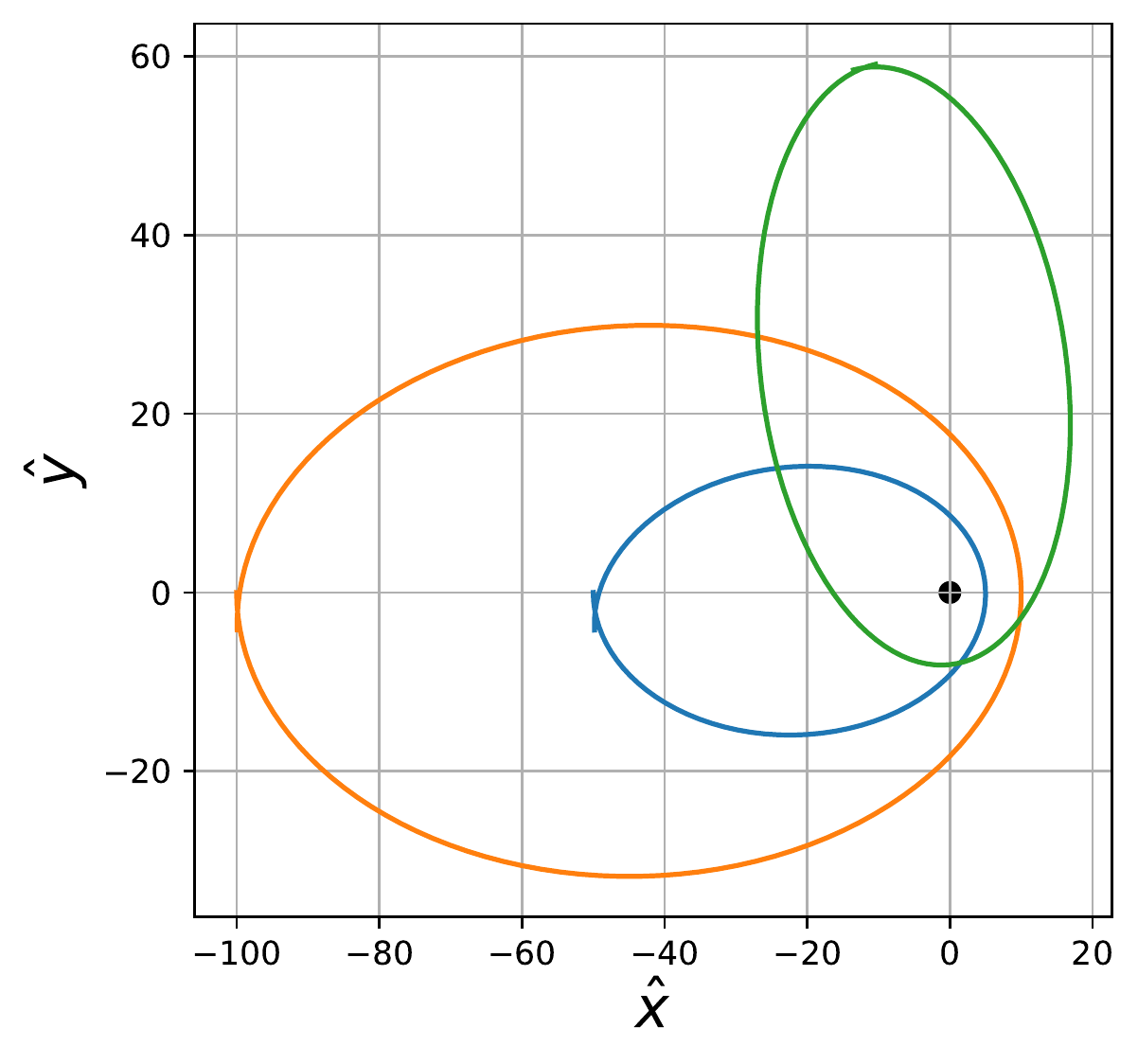}    \caption{Examples of elliptical orbits in the equatorial plane at different distances from the black hole. Orange corresponds to a distant orbit with $\hat{r}_1= 10$ and $\hat{r}_2 = 100$. Blue corresponds to an orbit close to $\hat{r}_1 = 5 $ and $\hat{r}_2= 50$. The green curve is the projection of an orbit with with the semimajor axis forming an angle of $\theta_0 = 80^{\circ}$  and $\hat{r}_1 = 8$ and $\hat{r}_2= 60$. The axis are described in terms of the dimensionless Cartesian coordinates $\hat{x}, \hat{y}$. }
    \label{fig:3-tray}
\end{figure}

We generate several trajectories from which we obtain the data to determine the angular momentum of the black hole, the mass, and the parameters of the trajectories as described above. Fig.~\ref{fig:3-tray} shows 3 illustrative examples of equatorial, near, far and initially inclined trajectories.
In this work, we will consider two cases, one for a black hole with the mass of Sagittarius $A^*$ \cite{Pei_ker_2020}, and another such that the Schwarzschild radius equals one astronomical unit. 
For each case we will analyze a trajectory of a body orbiting near the black hole and another orbiting far from it. In each example, we will use our procedure to determine the angular momentum of the black hole for three different values of $j$ such that, $j= 0$, $j = 0.1$ y $j = 0.95$. Furthermore, in our examples, we will consider that the fiducial distance, $R_0$ is one AU (Astronomical Unit), 
so that $T_0 = 1.580203\,\times\,10^{-5}\,{\rm years}$. 
It proved useful to use the following values for the speed of light and the gravitational constant: $c=6.3283\,\times\,10^4 \, \frac{\rm AU}{\rm year}, \, G=39.46\,\frac{\rm AU^3}{\rm year^2\,M_\odot}$.

\begin{figure}[ht]
    \centering
\includegraphics[scale=0.5]{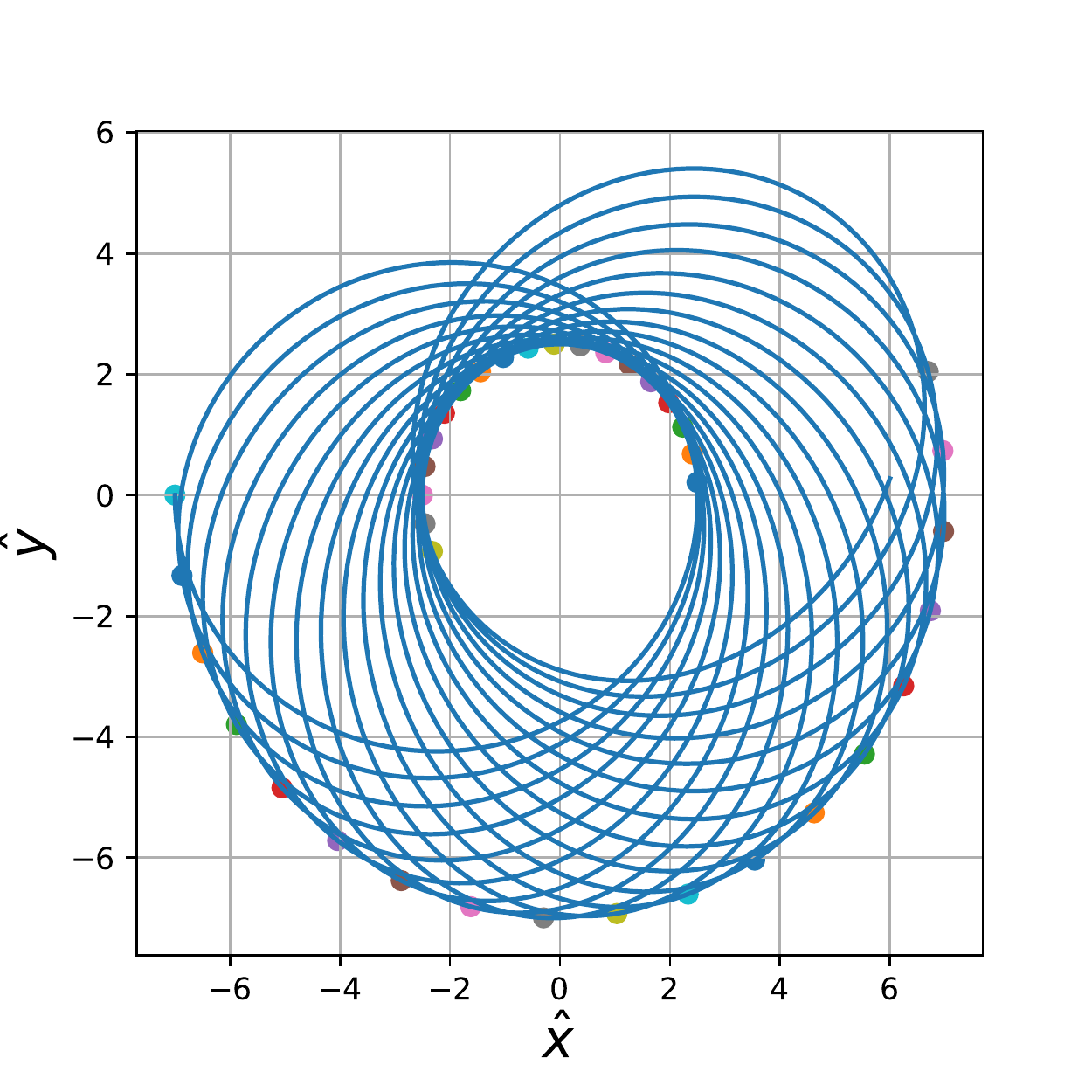}
    \caption{Trajectory around a black hole with mass $4.154\times 10^{6} \, M_\odot$ as in Sagittarius $A^*$. The orbit is equatorial and we have plot several turns. Marking in each turn the position of the apastron and periastron we observe the precession of these points clearly. The axis are described in terms of the dimension-less Cartesian coordinates $\hat{x}, \hat{y}$.}
    \label{fig:SgtA_cerca1}
\end{figure}

\begin{figure}
    \centering
    \includegraphics[scale=0.5]{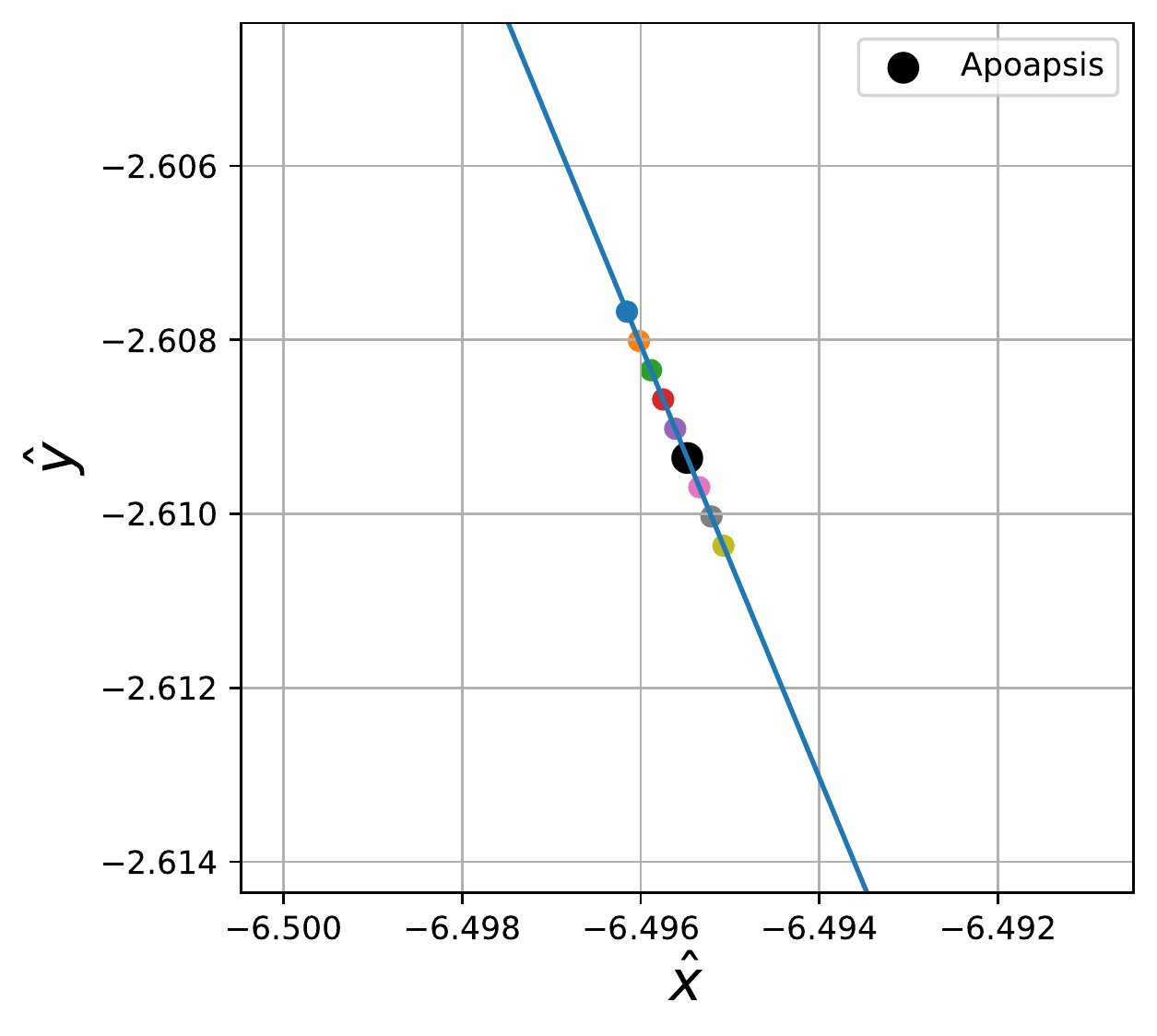}
    \caption{Close up of the trajectory of the body near the apoapsis. The colored dots are positions approaching and receding the apoapsis (black dot). It is presented a bounded orbit with $\hat{r}_1 = 2.5$, $\hat{r}_2 = 7$, and the axis are given in terms of the dimension-less Cartesian coordinates.}
    \label{fig:SgtA_cerca1_pos}
\end{figure}

In this way, in our first set of examples, the black hole has a mass of $4.154\,\times\,10^6\,M_\odot$ \cite{Pei_ker_2020}, and characteristic distance of one UA, that is $\sigma=0.20231343$. 
In Fig.~\ref{fig:SgtA_cerca1}, we present several periods of a trajectory considering $\hat{r}_1=2.5$ and $\hat{r}_2=7$. Both apsides precess in each orbit.
We can focus in any of the apsides, and choosing one in the apastron, for instance, we can mark the angular positions and times as the body moves near the chosen apsis. In Fig.~\ref{fig:SgtA_cerca1_pos} we present such observations, and in Tab.~\ref{tab:Ej_SgtrA_j0p95_cerca} we write down the values of the different angular positions and the corresponding times.
\begin{table}
    \centering
    \begin{tabular}{|c|c|c|}
    \hline
         Coordinate time   & Angular position  & Angular Velocity \\ 
         $\hat{t} \times 10^{2}$ & $\phi$ & $\hat{\Omega}\,\times\,10^{-3}$         \\ \hline
        6.5866275 & 12.9481042 & ~ \\ \hline
        6.5866923 & 12.9481559 & 7.9919264 \\ \hline
        6.5867571 & 12.9482077 & 7.9919244 \\ \hline
        6.5868219 & 12.9482595 & 7.9919226 \\ \hline
        6.5868867 & 12.9483113 & 7.9919207 \\ \hline
        6.5869515 & 12.9483630 & 7.9919189 \\ \hline
        6.5870162 & 12.9484148 & 7.9919171 \\ \hline
        6.5870810 & 12.9484666 & 7.9919154 \\ \hline
        6.5871458 & 12.9485184 & 7.9919137 \\ \hline
        6.5872106 & 12.9485701 & 7.9919121\\ \hline

    \end{tabular}
    \caption{Explicit values of angular and temporal positions of the body in the vecinity of the apoapsis, for the bounded orbit with $\hat{r}_1= 2.5$, $\hat{r}_2 = 7$.}
    \label{tab:Ej_SgtrA_j0p95_cerca}
\end{table}
Using the data presented in Tab.~\ref{tab:Ej_SgtrA_j0p95_cerca}, we can determine the angular difference between each successive points, $\Delta\,\phi|_{aps}$, as well as the difference between the corresponding coordinates times, $\Delta \, t|_{aps}$, first and second columns, so that we can compute the angular velocity between each pair:  $\hat{\Omega}|_{\hat{r}_{2i}}$, where $i$ stands for the chosen apsis. This result is presented in the third column for each pair. Taking the average of the obtained values we get that $\hat{\Omega}|_{\hat{r}_{2i}}=0.0079919$.

Using these values in Eq.~(\ref{eq:alg_j}), we 
obtain the following sixth degree algebraic equation for $j$:
\begin{eqnarray}
&-1.624204\,10^{-17}\,j^6 -8.292995\,10^{-12}\,j^5  -1.032902\,10^{-6}\,j^4 & \nonumber \\ 
&+ 6.559439\,10^{-3}\,j^3 - 8.874919\,j^2 + 193.5702\,j-176.0276=0,&    
\label{angularmoment}
\end{eqnarray}
we solve directly the system of algebraic equations, using the Maple software, and finally obtain the value of the black hole's angular momentum:
\begin{equation}
 j=0.9507917. 
\end{equation}

Finally, using these values in Eqs.~(\ref{eq:ENK_ecu}, \ref{eq:LfK_ecu}), we obtain the corresponding values for the energy $\hat{E}_N$ and azimuthal momentum of the observed body:
\begin{eqnarray}
\hat{E}_N&=&-4.277410\,\times\,10^{-3}, \\
\hat{L}_\phi&=&0.3940907,
\end{eqnarray}
so that we completely determine the trajectory parameters as well as the black hole angular momenta.

It is interesting to note that the Eq. \eqref{angularmoment} for $j$  gives other allowable values for the black hole's angular momentum, we obtain two complex values, but also the values $j=21.20694,\,1939.205,\,4152.716,$ $-2.582786\, \times\,10^5, -2.584237\,\times\,10^5$, which would imply that within the specific known trajectory parameters, the black hole at Sagittarius A* could also correspond to a naked singularity (hypothetical gravitational singularity without an event horizon \cite{Hawking:1970zqf}) with the same mass. We will see that using other other trajectories, the naked singularities can be discarded and only the physical solution remains.

The actual values of the parameters in the code where $j=0.95$, and $\hat{E}_N=-4.27740471\,\times\,10^{-3}, \, \hat{L}_\phi=0.39409197$, so that the values obtained, have a difference with respect to the actual values $ |\Delta\,j|=8.34\,\times\,10^{-3}\,\%,  \, |\Delta\,\hat{E}_N|=1\,\times\,10^{-4}\,\%, \, \Delta\,\hat{L}_\phi=3.1\,\times\,10^{-5}\,\%$.

\begin{figure}[ht]
    \centering
    \includegraphics[scale=0.5]{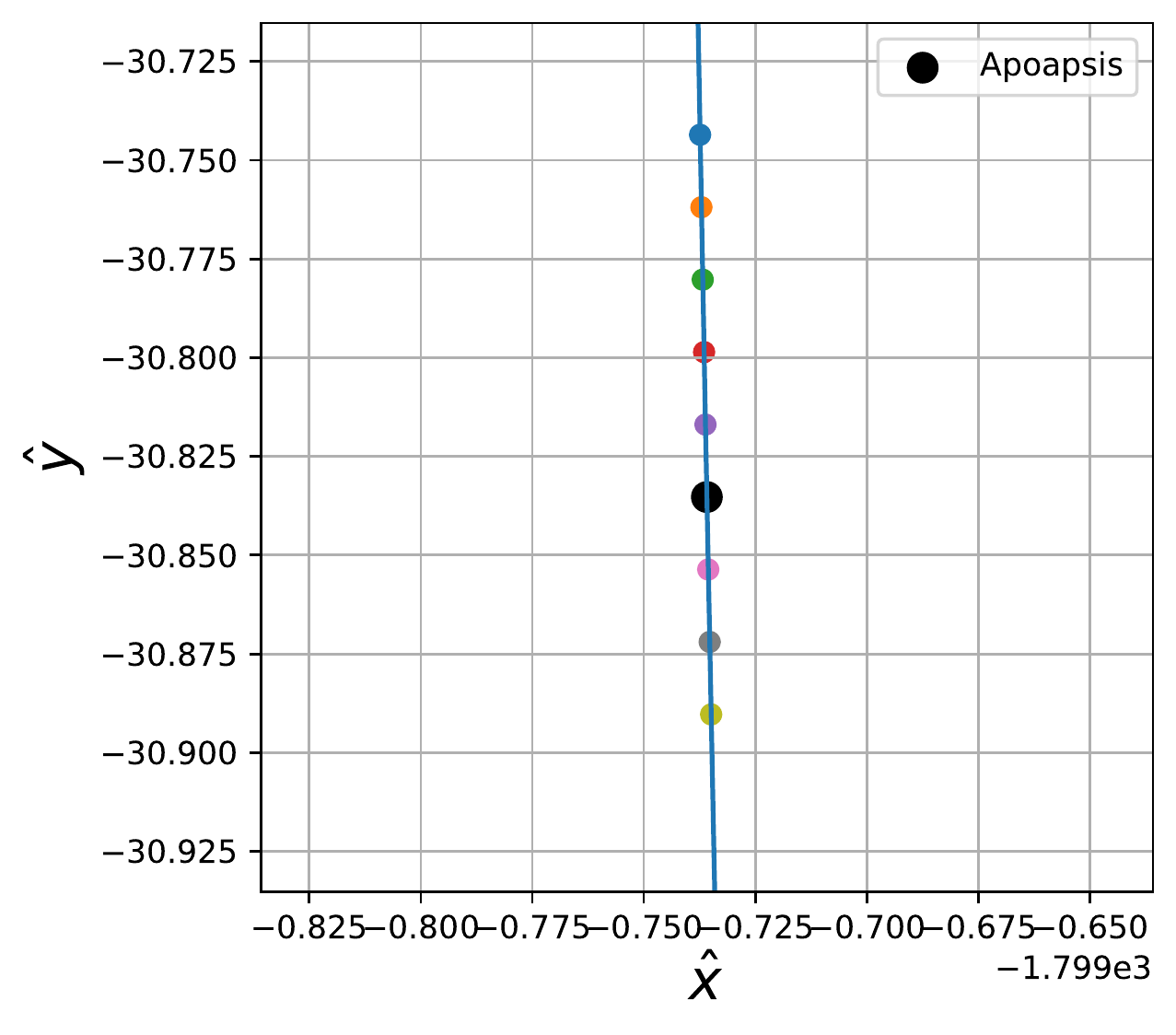}
    \caption{Close up of the trajectory of the body near the apoapsis. The colored dots are positions approaching and receding the apoapsis (black dot). It is presented a bounded orbit with $\hat{r}_1 = 45$, $\hat{r}_2 = 1800$, and the axis are given in terms of the dimension-less Cartesian coordinates.}
    \label{fig:SgtA_lejos1}
\end{figure}

We continue presenting the position data for another body orbiting the same black hole in another trajectory, also equatorial, but further away from the black hole, with apsides at $\hat{r}_1=45, \hat{r}_2=1800$, see Fig. \ref{fig:SgtA_lejos1}.
Again, we can measure the angular positions and the corresponding coordinates times for the several events and obtain the average value for the angular velocity of $\hat{\Omega}|_{\hat{r}_{2i}}=5.8561054\,\times\,10^{-7}$, Tab. \eqref{tab:Ej_SgtrA_j0p95_lejos}.
\begin{table}
    \centering
    \begin{tabular}{|c|c|c|}
    \hline
         Coodinate time   & Angular position  & Angular Velocity \\ 
         $\hat{t} \times 10^{6}$ & $\phi$ & $\hat{\Omega}\,\times\,10^{-7}$         \\ \hline
        1.7404791 & 12.5834512 &  \\ \hline
        1.7404965 & 12.5834614 & 5.8561057 \\ \hline
        1.7405139 & 12.5834715 & 5.8561056 \\ \hline
        1.7405313 & 12.5834817 & 5.8561055 \\ \hline
        1.7405487 & 12.5834919 & 5.8561054 \\ \hline
        1.7405661 & 12.5835021 & 5.8561054 \\ \hline
        1.7405835 & 12.5835123 & 5.8561053 \\ \hline
        1.7406009 & 12.5835225 & 5.8561053 \\ \hline
        1.7406183 & 12.5835327 & 5.8561053 \\ \hline
        1.7406357 & 12.5835429 & 5.8561054 \\ \hline
    \end{tabular}
    \caption{Explicit values of angular and temporal positions of the body in the vecinity of the apoapsis, for the bounded orbit with $\hat{r}_1 = 45$, $\hat{r}_2 = 1800$.}
    \label{tab:Ej_SgtrA_j0p95_lejos}
\end{table}
Using these values of the apsis, $\sigma$ and $\hat{\Omega}$ in Eq.~(\ref{eq:alg_j}), we now obtain %
\begin{eqnarray}
&&-3.105015\,10^{-24}\,j^6 -5.696254\,10^{-12}\,j^5 -2.612253\,j^4 \nonumber\\
&&+ 2.179947\,10^8\,j^3 -4.544723\,10^{15}\,j^2+ 4.216119\,10^{17}\,j \nonumber\\ 
&&-3.962144\,10^{17}=0, 
\end{eqnarray}
with solution:
\begin{equation}
j=0.949478662400226, 
\end{equation}
which is a very close value to the one obtained with the first trajectory, as it should as long as it is the same central body. For the energy and momenta of the body we now obtain:
\begin{eqnarray}
\hat{E}_N&=&-2.21837296234096\,\times\,10^{-5}, \\
\hat{L}_\phi&=&1.89742061339787,
\end{eqnarray}
again completely determining the trajectory parameters as well as the black hole angular momentum. It is interesting to mention that the other values obtained for $j$ in this case, which correspond to a naked singularity cases, do not coincide with the naked singularity solutions obtained in the previous case for a closer trajectory; the only value that coincides is the one in the interval between zero an one, thus showing that the algorithm gives only one consistent value when using several trajectories.  Finally, regarding the errors with respect the input values in this example, we have that 
$ \Delta\,j=5.48\,\times\,10^{-2}\,\%, \, \Delta\,\hat{E}_N=9.81\,\times\,10^{-8}\,\%, \, \Delta\,\hat{L}_\phi=-2.009\,\times\,10^{-6}\,\%$. Again showing a good performance of our procedure in the case of mock data describing a trajectory with apsides far from the black hole.

We perform another two similar experiments for a black hole of the same mass but with different angular momentum. In each case we use the same values of the apsis at the closer and farther away case. We repeat the series of experiments for a black hole with different mass,  $M=5.074427\,\times\,10^7\,M_\odot$, and we again take the characteristic distance, $R_0$ as one AU, so that the fiducial distance is given by the Schwarzschild radius. We call this object the AU Black Hole. 
\begin{table*}
    \centering
    \scriptsize{
    \begin{tabular}{|l|l|l|l|l|l|l|l|l|}
    \hline
        Name & $\hat r_1$ & $ \hat r_2$ &  $ \hat{\Omega} = \frac{d\phi}{dt}$ & j & $E_n$ & $L_\phi$ & $\Delta Error$\\ \hline
        SgtrA   & 2.5 & 7    & 8.01080007$\times 10^{-3}$ & -3.7910667$\times 10^{-6}$ & -4.2699918$\times 10^{-3}$ & 0.3954743 & 0.0003791\\ \hline
                & 45  & 1800 & 5.85631667$\times 10^{-7}$ & -4.5293004$\times 10^{-4}$ & -2.2183689$\times 10^{-5}$ & 1.8974908 & 0.0452930\\ \hline
                & 2.5 & 7    & 8.00872934$\times 10^{-3}$ & 0.09999440 & -4.2708104$\times 10^{-3}$ & 0.3953235 & 0.0055974\\ \hline
                & 45  & 1800 & 5.85629423$\times 10^{-7}$ & 0.09953801 & -2.2183693$\times 10^{-5}$ & 1.8974833 & 0.4619899\\ \hline
                & 2.5 & 7    & 7.99191908$\times 10^{-3}$ & 0.95092447 & -4.2774040$\times 10^{-3}$ & 0.3940916 & 0.0973130\\ \hline
                & 45  & 1800 & 5.85610549$\times 10^{-7}$ & 0.95171333 & -2.2183729$\times 10^{-5}$ & 1.8974206 & 0.1803511\\ \hline\hline
        AUBH    & 10  & 15   & 1.10863915$\times 10^{-2}$ & 1.64157790$\times 10^{-6}$ & -1.9083968$\times 10^{-2}$ &   2.6211119 & 0.0001641\\ \hline
                & 500 & 2500 & -3.2686032$\times 10^{-6}$ & 3.83257318$\times 10^{-4}$ & -1.6661098$\times 10^{-4}$ & -20.4335397 & 0.0383257 \\ \hline
                & 10  & 15   & 1.10510497$\times 10^{-2}$ & 0.2500000 & -1.91792180$\times 10^{-2}$ &  2.6036784 & 0.0000164 \\ \hline
                & 500 & 2500 & -3.2686367$\times 10^{-6}$ & 0.2503675 & -1.66610298$\times 10^{-4}$ &-20.4337992 & 0.1470335 \\ \hline
                & 10  & 15   & 1.09559305$\times 10^{-2}$ & 0.9699999 & -1.94221295$\times 10^{-2}$ &  2.5573103 & 4.91$\times 10^{-11}$ \\ \hline
                & 500 & 2500 & -3.2687342$\times 10^{-6}$ & 0.9701330 & -1.66608303$\times 10^{-4}$ &-20.4345537 & -0.013718  \\ \hline
    \end{tabular}
    }
    \caption{Results of the analysis of the orbits around a hole of mass $n=4.154\times 10^6$ type Sagittarius $A^\ast$ and for $n = 5.074427\times10^7$ solar type. The table shows the calculated j and its corresponding error.}
\end{table*}

%
\section{Comments on non equatorial trajectories}
\label{sec:Non_equatorial}
We have shown that our procedure, which asumes that the apsides of a bounded trajectory are known and using the measurements of the angular displacement for equatorial trajectories, allows us to determine the angular momentum of the black hole, as well as the parameters of the trajectory. For the trajectories moving out of the equatorial plane, as is shown in the right panel of Fig.~\ref{fig:traj_j0p98}, the apsides, although they occur at fixed radius of the apastron and periastron, $\hat{r}_2, \hat{r}_1$, have a precession in both angles $\theta$ and $\phi$, so that, in general, we cannot impose conditions on the respective angular velocities, $u^\theta, u^\phi$, and the measured angular displacement has to be related with changes in both angular coordinates.
There are, though, some events with singular characteristics, like the supreme value of the $\theta$-angle of the apastron (infimum for the periastron) where the angular velocity $u^\theta$ is equal to zero, as well as other resonant points \cite{Brink:2015roa} where our procedure can be applied by means of the use Eq.~(\ref{eq:uth}) to express the Carter's constant, say, in terms of the other unknowns and then a procedure similar to the one used in the equatorial case can be followed and thus determine the value of the black hole's angular momentum and the orbital parameters based on the observations of the motion. However, it is not possible to determine, with the observation of a single cycle of a trajectory, whether or not it is the one corresponding to the supreme angle and, in general, it will no be so, and we face similar problems when resonances are considered.
For a general, non-equatorial trajectory, one can assume that the angular and time displacements of the orbiting body near an apsis, $\Delta\,\alpha, \, \Delta\,t$, can be determined, but it is needed to define a way to relate such temporal and angular displacement near the apsides,  with the corresponding angular displacements of the angles, and then, with the angular velocities, so that one would be able to derive relations of the type shown in the present procedure and thus determine the angular momentum of the black hole.
We have seen in our experiments that the tangential velocity at the apsis, defined as $v_t=\sqrt{g_{\theta\theta}\,{u^\theta}^2 + g_{\phi\phi}\,{u^\phi}^2}|_{ap}$, has a constant value at each of the afeasers, and another constant value at each of the periasters. This is an interesting fact that could be useful in a procedure to determine the angular momentum of the black hole from observations of trajectories laying outside the equatorial plane. Ideas in this direction are currently under way and will be presented in a forthcoming work.

\section{Final remarks}
\label{sec:final_comms}

We have presented a procedure where, by means of the determination of the angular displacements of a body moving in a Kerr background in a equatorial bounded trajectory, it is possible to determine the angular momentum of the central black hole. 
Our procedure is not a perturbation method, so that there is no need to make assumptions on the magnitude of the black hole's angular momentum, nor on the moving body distance to the black hole. We have seen in the examples presented that,  using two trajectories, our procedure could also be used to single out the actual value of the black hole's angular momentum. From a given observation, other values of $j$ are solutions to the algebraic equation (\ref{eq:alg_j}), several of them correspond to naked singularities. However, when using a second trajectory, only one value coincides (in our case, the value which was given as input in the code) so that all the other values, implying naked singularities, can be discarded.
We have also derived an expression for determining the mass of the black hole, in the case when it is a Schwarzschild background, or in the case within the Keplerian description, using the measurement of the tangential velocity at one of the apsis. In appendix \ref{sec:BH_Mass}, we test such expression for the mass in terms of the tangential velocity at the apsis, Eqs.~(\ref{eq:sigma_Schw}, \ref{eq:sigma_K}), in a Kerr background and obtain that for equatorial orbits, the expression for the mass can be accurate to the millionth. We have also tested these expressions even for the case when the trajectory is not equatorial in the Kerr background and, using Eqs.~(\ref{eq:sigma_Schw}, \ref{eq:sigma_K}) inserting the corresponding angular displacement as if the motion took place in the equatorial plane, obtain also excellent results. In this Appendix, we included also the determination of the mass using directly Kepler's third law, Eq.~(\ref{eq:sigma_3aLK}), where we defined the period as the time it takes to return to a given radial value, independently of the angular values, and obtain that Kepler's third law actually gives a very good approximation for the mass, even when considering data from trajectories passing near the Kerr black hole horizon radius.  
Finally, we presented some ideas on how to extend these results to the case of non-equatorial obits; it is an on-going work, and we expect to present results in the near future. Another step forward of our procedure is to use actual observational data, and it is needed to work with the precision with which the data is obtained, for instance, with the positions near the apsides of the observed stars orbiting around  Sagittarius A$^*$, for a trajectory that lays in the equatorial plane, as the ones presented in \cite{Gravity_2021b, Ghez:2008ms}. Also, observations of the accretion disks, if they are on the equatorial plane and the ISCO (innermost stable circular orbit) orbit could be identified with the geodesic motion of a body at that radius and apply our procedure.


The determination of the angular momentum of astrophysical black holes will provide an important basis for testing the predictions of the General Theory of Relativity and for an understanding of the characteristics of the central region of our Galaxy.

\appendix
\section{Estimation of the mass and angular momentum of black holes} \label{sec:Estimation}

\subsection{Mass determination} \label{sec:BH_Mass_g}

The most accurate and reliable mass determination methods in astronomy are dynamical. Newton's and Kepler's laws are used to explain the observed movements of orbiting objects.  In the ideal scenario of a massive black hole, like the one in the center of our own Galaxy, one can measure the orbital parameters of stars  around it and calculate its mass, as for example in \cite{Ghez2005}, see also Appendix \ref{sec:BH_Mass}.

Since dynamical methods need spatially resolved data, they can be applied only to very near objects, like Sagittarius A*, or in some not-so-close galaxies, but which have special characteristics like M87, but in general the dynamical methods are not applicable.

In astronomy it is usual to distinguish between two ways to estimate the masses of gigantic black holes. Primary methods: These are direct measurements like the one used with Sagittarius A*.
Usually a primary method is based on how stellar velocities are affected by the gravitational force from a distributed mass in the nucleus (stars or gas) and by the black hole itself. The principal idea is to measure those velocities via Doppler shifts or proper motions (in the nearest galaxies), or Doppler broadening of stellar absorption lines in the integrated spectra (in galaxies beyond the local group). Also, ionized gas can be present and the estimated mass inside it can be calculated from Doppler shifts in the emission lines. The most popular  methods include stellar and gas dynamical modeling and reverberation mapping  \cite{Blandford,Peterson,Shen}. This technique allow us to measure super massive black hole masses in bright AGNs \footnote{Active Galactic Nuclei (AGN), are a subset of galaxies which are characterized by their very powerful bright in the central regions, this emission is believed to be powered by a supermassive black hole which accretes matter to a disk around it and and during the process, gravitational energy is converted to electromagnetic radiation}. In this case masses are determined through time resolution of the AGN variability, rather than spatial resolution. It measures the time delay between changes in the continuum emission (likely arising from the accretion disk) and the response to those changes in the broad emission lines (arising from the photoionized Broad Line Region (BLR)\footnote{In the standard model of AGNs, there are different regions around the center and the light we detect, comes not from the accretion disk itself but from a more distant region surrounding it. Photons are generated in the inner regions and in the disk and  are dispersed in the so called Broad Line region, which is not so close to the disk.}).

In this sense, the AGN Black Hole Database is a very useful compilation of spectroscopic reverberation mapping studies with black hole masses of over more than 85 galaxies with redshift $z<0.4$ \cite{Bentz2015}. 

Secondary methods: They are usually based on empirical correlations and are calibrated using primary mass estimates. These so called {\it scaling relations}, which were initially observed some decades ago, show  linear correlations between the mass of the black hole ($M_{\bullet}$) and the luminosity of the galactic bulge (in the optical band, $ L_V$) and also with the stellar velocity dispersion ($M_{\bullet}-L_V$  and $M_{\bullet}-v_s$) in elliptic galaxies \cite{Magorrian, Ferrarese, Gebhardt}. Most of the actual estimates for super massive black hole masses come from a combination of techniques and are continuously under revision.

Many of the most recent variants of spectroscopic methods include different {\it mass estimators} which use different broad line emission lines like ${\rm H}{\alpha}, \, {\rm H}{\beta}, \, {\rm MgII}\, \lambda 2798, \, {\rm CIV}\, \lambda \, 1549, {\rm Fe\,III}$ and others \cite{Mediavilla, DallaBonta}. 

Scaling relations have become a preferred way to estimate black hole masses of distant active galaxies and quasars because of the ease with which the method can be applied to large samples. However, most of the spectra available have two characteristics which must be considered in the analysis. They are single epoch and there is only one spectra per galaxy (usually all the light emitted by the galaxy is considered in the spectra, not only that coming from the central region).

However, with the use of IFUs techniques\footnote{IFU (Integral Field Unit) spectroscopy is an observing technique that provides spectral information over a 2D field of view for a large number of spatial elements (spaxels), each spaxel covers a different region of the target galaxy. The resulting spectra are arranged into a datacube which contains  the entire 2D field plus the third dimension drawn from the spectrograph. Depending on the type of IFU being used, up to hundreds or thousands of spectra are recorded simultaneously in any single exposure.} the situation is now much better, because the availability of detailed spectra for different regions of the galaxies has improved the confidence in the mass estimations (at least for galaxies in the nearby universe, $z<0.1$). So far, the use of high resolution spectra for different regions of a galaxy provides precise values of the stellar velocity dispersion,  they are considered highly reliable and then the use of scaling relations in the estimation of the black hole mass becomes reliable as well.

In those cases where spectra are not available and the characteristics of the galaxy (distance, inclination etc.) do not allow the use of these methods, we use photometry  \cite{lauer, rusli}. This is also a secondary method which uses the light profile of the galaxy to estimate the mass of the central object. The advantage of these photometric techniques is that the number of galaxies with photometric data is far greater than those with spectroscopic data. 

In any case, the accuracy of the estimations ultimately depends on the good calibration of models with observed data. Recently an attempt to combine both techniques has been made and it has shown promising results \cite{Romero-Cruz}. It has been proposed the construction of an adjusting factor to improve the photometric mass estimations, it was calibrated using IFU spectroscopic data from the CALIFA survey\footnote{ CALIFA (The Calar Alto Legacy Integral Field Area Survey) observes a statistically well-defined sample of $\sim$ 600 galaxies in the local universe using an Integral Field Spectrophotometer, mounted on the Calar Alto 3.5 m telescope, in Spain.} and enhanced scaling relations  \cite{mcconnell}. The first stage of the model has been developed for nearby elliptic galaxies ($z<0.03$), and it is planned to be extended to galaxies at any distance.

With respect to black holes inside our own galaxy, the techniques are somewhat different since so far we can only infer the presence of a black hole of a few solar masses in binary systems by means of the X-ray emission that can be detected coming from the accretion disk. Intermediate mass black holes (of the order of $100$ to $\sim 10^4-10^5\, M_{\odot}$) continue to be elusive \cite{greene}.

Different techniques to estimate the mass of black holes are based mainly on observations of the $z$-shifted light emitted by the bodies moving in its vicinity and with a Keplerian model, which as it is shown in Appendix \ref{sec:BH_Mass}, is a good approximation; although the other expressions presented, specially Eq. (\ref{eq:sigma_Schw}), offer a more accurate value. In any case, based on this brief review, we can consider that the mass of a given black hole can be estimated. Also, it is important to mention that in our proposal to determine the black hole's angular momentum, we consider that the mass of the black hole is known.    

\subsection{Angular momentum determination} \label{sec:BH_j_g}

There are not as many works regarding the determination of the angular momentum as there are for the mass. However, several interesting proposals have been presented and we will mention some of them:

We start mentioning the method proposed  by Hioki and Maeda \cite{Hioki:2009na}, where they determined the rotation parameter and the inclination angle by means of observations of the shape of the shadow, which is a function of these two parameters. Defining observable quantities characterizing the apparent shape (its radius and distortion parameter) they find that the spin parameter and the angle of inclination can be determined by them, and thus,  by observation of the distortion parameter, the mass of the black hole and the inclination angle of a trajectory, they can determine the spin parameter. Another proposal in this direction is presented by \cite{Li:2013jra}, where the authors considered the measurement of the Kerr spin parameter of the Bardeen and Hayward regular black holes from their shadow and then compared the result with the estimate inferred from the K$\alpha$ iron line and from the frequency of the innermost stable circular orbit.

In \cite{Brink:2015roa} it was studied a complete characterization of the location of resonant orbits in the Kerr spacetime for all possible black hole spins and orbital parameter values. The authors have defined a resonant orbit as a geodesic for which the longitudinal and radial orbital frequencies are commensurate. Resonant effects may have observable implications for the in-spirals of compact objects into a super-massive black hole. At these locations rapid changes in the orbital parameters could produce a measurable phase shift in the emitted gravitational and electromagnetic radiation. Resonant orbits may also capture gas or larger objects, leading to further observable characteristic electromagnetic emission. 
Indeed, the concentration of low order resonances near the black hole and their absence further out has implications for testing the no-hair theorems using a super-massive black hole such Sagittarius A$^*$. 

The no-hair theorems state that, provided the cosmic censorship and causality axioms hold, if the black hole’s mass and spin are known the quadrupole moment is fixed \cite{Will:2007pp}. Indeed, in a seminal work, C. Will presented a procedure to compare the quadrupolar term in a PPN expansion, with terms involving the ratio of the squared angular momentum to the mass of the Kerr black hole. The quadropolar momentum can be determined by means of the observations of the bodies moving in the compact object near region and thus be used as a test for the validity of the theory. Following Will's idea, the authors in \cite{Liu_2012}  have shown that recording the time of arrival signals coming from a pulsar that would be orbiting Sagittarius A$^\ast$, with orbital period $\sim$ 4 months, and for observations during several years with the Square Kilometer array (SKA), will be possible to measure the mass of Sagittarius A$^\ast$ to a precision of $10^{-6}$ , the spin $10^{-3}$ and the quadrupole moment to $10^{-2}$. Such measurements, as the one presented in this manuscript, allow to have a spacetime background determined with large accuracy and thus provide a test of the no-hair theorems. The detection of a pulsar even closer to the central object could allow the extraction of additional multipole moments through shorter term monitoring, thus mapping out more details of the structure of the central black hole.

In \cite{Gammie:2003qi} the authors have considered astrophysical processes that influence the spin evolution of black holes and present an interesting analysis showing the intervals for the final value of the black holes's angular momentum in several scenarios of formation. In \cite{Kulkarni_2011}, the authors  determine the black hole's angular momentum using the X-ray spectra of accretion discs of eight stellar-mass black holes with the thermal continuum fitting method.
Their method applies for razor-thin discs however, in their simulations they show that the values obtained for more general cases are in good agreement with other methods.

It is interesting to mention the proposal described in \cite{Herrera-Aguilar:2015kea}, where the authors obtained an analytic expression for the mass and angular momentum of a Kerr black hole in terms of the redshift and blue shift of photons emitted by stars in stable equatorial and circular orbits around the black hole.

And related to this idea, in \cite{Vitale_2014} the authors analyzed simulated signals emitted by spinning binaries with several values of masses, spins, orientation, and signal-to-noise ratio, as detected by an advanced LIGO-Virgo network. They found that for moderate or high signal-to-noise ratio the spin magnitudes can be estimated with errors of a few percent (5-30)  for  neutron star - black hole (black hole - black hole) systems. 

Finally, bounds have been established on the rotation parameter by considering the Lense-Thirring effect on the distribution of the S-stars  \cite{Fragione:2020khu}, obtaining the constraint $j\lesssim 0.1$. Also within the Event Horizon Telescope first results \cite{EventHorizonTelescope:2022urf, EventHorizonTelescope:2022xqj} limits have also been established for $j$ and for the angle between the line of sight and the spin axis of Sagittarius A*.

In this way, we can see that there are several interesting proposals, some of which are still at a theoretical level. In the next section, we present our proposal which, as mentioned above, is based on observations of the positions of the stars near the apsis.

\section{Equations of motion and conserved quantities} \label{Sec:Eqs-motion}

Starting from the Hamiltonian $\hat{{\cal H}}=\frac{1}{2} \hat{p}_\mu \hat{p}^\mu -\frac{1}{2}$
:
with the Hamilton's equations:
$\dot{\hat{q}}^\mu=\frac{\partial\,{\cal H}}{\partial \hat{p}_\mu}, \dot{\hat{p}}_\mu=-\frac{\partial\,{\cal H}}{\partial \hat{q}^\mu}$ we obtain the following dynamical equations:
\small{
\begin{eqnarray}
\frac{d\,\hat{t}}{d\,\hat{\tau}}&=&\frac{\Delta_{1+}\,\sqrt{1+2\,\hat{E}_N}+2\,\frac{\sigma^2}{\hat{r}}\,\hat{p}_r}{\Delta_{0+}}, \label{eq:ham_t}\\
\frac{d\,\hat{r}}{d\,\hat{\tau}}&=&\frac{-2\,\frac{\sigma^2}{\hat{r}}\,\sqrt{1+2\,\hat{E}_N} + \Delta\,\hat{p}_r + \frac{j\,\sigma^2}{\hat{r}^2}\,\hat{L}_\phi}{\Delta_{0+}}, \label{eq:ham_r}\\
\frac{d\,\theta}{d\,\hat{\tau}}&=&\frac{\hat{L}_\theta}{\hat{r}^2\,\Delta_{0+}}, \label{eq:ham_th}\\
\frac{d\,\phi}{d\,\hat{\tau}}&=&\frac{j\,\sigma^2\,\hat{p}_r + \frac{\hat{L}_\phi}{\sin^2\theta}}{\hat{r}^2\,\Delta_{0+}}, \label{eq:ham_f}\\
\frac{d\,\hat{p}_r}{d\,\hat{\tau}}&=&\frac{\Delta_{0-}\,\sigma^2}{{\hat{r}}^2\,\Delta_{0+}}\,\left(1+2\,\hat{E}_N + {\hat{p}_r}^2 + 2\,\sqrt{1+2\,\hat{E}_N}\,\hat{p}_r\right) - 2\,\frac{\sigma^2\,j}{\hat{r}^3\,\Delta_{0+}}\,\hat{p}_r\,\hat{L}_\phi \nonumber\\
&&-\frac{\sigma^4\,j^2}{{\hat{r}}^3\,\Delta_{0+}}\,\left(2\,\cos^2\theta\,\hat{E}_N + {\hat{p}_r}^2\,\sin^2\theta\right)  - \frac{\hat{C}^2+{\hat{L}_\phi}^2}{\hat{r}^3\,\Delta_{0+}}, \label{eq:ham_pr}\\
\frac{d\,\hat{L}_\theta}{d\,\hat{\tau}}&=&-\frac{\cos\theta}{\hat{r}^2\,{\Delta_{0+}}^2\,\sin^3\theta}\,{\hat{L}_\phi}^2 + - 
2\,\frac{\sigma^2\,j\,\hat{p}_r\,\hat{L}_\phi}{\hat{r}^2} \\ &&\frac{j^2\,\sigma^4\,\cos\theta\,\sin\theta}{\hat{r}^2\,{\Delta_{0+}}^2}\,\left[\Delta\,{\hat{p}_r}^2 + \frac{\hat{C}^2 + \frac{\left(1-3\,\cos^2\theta+\cos^4\theta\right)\,{\hat{L}_\phi}^2}{\sin^4\theta}}{\hat{r}^2}
+ \right. \nonumber \\ 
&&
\left. - 
2\,\frac{\sigma^2}{\hat{r}}\,\left(1 + \left(2 - \frac{\sigma^2\,j^2\,\cos^2\theta}{\hat{r}}\right)\,\hat{E}_N + 2\,\sqrt{1+2\,\hat{E}_N}\,\hat{p}_r\right)
 \right]. \nonumber
 \label{eq:ham_Lth}
\end{eqnarray}
}
where we have defined
\begin{eqnarray}
\Delta&=&1 - \frac{2\,\sigma^2}{\hat{r}} + \frac{\sigma^4\,j^2}{\hat{r}^2}, 
\nonumber \\
\Delta_{1\pm}&=&1 \pm 2\,\frac{\sigma^2}{\hat{r}} +  \frac{\sigma^4\,j^2 \cos^2\theta}{\hat{r}^2}, 
\end{eqnarray}
$\Delta_{0\pm}$ are defined above, Eq.~(\ref{Delta0}), and we have written the Hamilton equations in terms of the conserved quantities: $\hat{L}_\phi:=\frac{\partial\,{\cal L}}{\partial\,\dot{\phi}}$ is the conserved quantity associated with the azimuthal symmetry and, $\frac{\partial\,{\cal L}}{\partial\,\dot{\theta}}$, as is described in the Appendix \ref{sec:Carter}, the right ascension momenta can be written in terms of $\hat{C}$ as:
\begin{equation}
\hat{L}_\theta=\sqrt{\hat{C}^2 - \cot^2\theta\,{\hat{L}_\phi}^2 - 2\,j^2\sigma^4\,\hat{E}_N\,\cos^2\theta}. \label{eq:Lth}    
\end{equation}
The conserved quantity associated with the time symmetry, $\hat{p}_0:=\frac{\partial\,{\cal L}}{\partial\,\dot{x^0}}$, has been related to the Newtonian energy, $\hat{E}_N$ in the following way: Using the normalization of the four momenta, $g^{\alpha\beta}\hat{p}_\alpha\hat{p}_\beta=-1$, with $p_\alpha=m\,c\,\hat{p}_\alpha$, being $m$ the mass of the particle, one can express the radial component of the momenta $\hat{p}_r$ in terms of the conserved quantities $\hat{p}_0, \hat{L}_\phi, \hat{C}$ and, substituting in the Hamilton's equation $\hat{u}^r=\frac{\partial\,{\cal H}}{\partial\,\hat{p}_r}$, one can finally take the Keplerian case and thus identify
\begin{equation}
\hat{p}_0=-\sqrt{1+2\,\hat{E}_N}. \label{eq:p0-EN}
\end{equation}

\section{Carter's constant derivation} \label{sec:Carter}

As it is well known, the angular momentum is not a conserved quantity in the Kerr spacetime, nor is its magnitude. This fact implies that the motion, in general, will no remain in a plane, but precesses respect the equatorial plane. This is a very characteristic feature of the Kerr spacetime and will be discuss in detail in the next appendix. For now, let us analyze the consequences that the magnitude of the total angular momentum, $L^2$ is not a conserved quantity.

Indeed, from the usual definition of the total 
angular momentum in terms of the angular momentum components:
%
\begin{eqnarray}
\hat{L}^2&=&{\hat{L}_\theta}^2 + \frac{{\hat{L}_\phi}^2}{\sin^2\theta}, 
\\
&=&\hat{r}^4\,{\Delta_{0+}}^2\,{\hat{u}^\theta}{}^2 + \frac{{\hat{L}_\phi}^2}{\sin^2\theta}, 
\end{eqnarray}
%
where we are using the dimensionless expressions. It is interesting to explore the amount that this quantity is not a conserved quantity:
\begin{eqnarray}
\frac{d\,\hat{L}^2}{d\,\hat{\tau}}&=&2\,{\hat{L}_\theta}\,\frac{d\,\hat{L}_\theta}{d\,\hat{\tau}} - 2\,\frac{\cos\theta}{\sin^3\theta}\,{\hat{L}_\phi}^2\,\frac{d\,\theta}{d\,\hat{\tau}}, \\
&=&2\,{\hat{L}_\theta}\,\frac{\partial\,{\cal{L}}}{\partial\,\theta} - 2\,\frac{\cos\theta}{\sin^3\theta}\,{\hat{L}_\phi}^2\,\hat{u}^\theta, \\
&=&2\,{\hat{L}_\theta}\,\left(\frac{{\hat{L}_\phi}^2\,\cos\theta}{\hat{r}^2\,\Delta_{0+}\sin^3\theta} - \frac{\left(\frac{k}{\sigma^2} + {\hat{p}_0}^2\,\sigma^2\right)\,j^2\,\sigma^4\,\cos\theta\,\sin\theta}{\hat{r}^2\,\Delta_{0+}}\right)\nonumber\\ 
&& - 2\,\frac{\cos\theta}{\sin^3\theta}\,{\hat{L}_\phi}^2\,
\left(\frac{\hat{L}_\theta}{\hat{r}^2\,\Delta_{0+}}\right), \\
&=&-2\,\frac{\left(\frac{k}{\sigma^2} + {\hat{p}_0}^2\,\sigma^2\right)\,\hat{L}_\theta\,j^2\,\sigma^4\,\cos\theta\,\sin\theta}{\hat{r}^2\,\Delta_{0+}}, \\
&=&-2\,\left(\frac{k}{\sigma^2} + {\hat{p}_0}^2\,\sigma^2\right)\,\hat{u}^\theta\,j^2\,\sigma^4\,\cos\theta\,\sin\theta, \\
&=&\frac{d\,\left(\left(\frac{k}{\sigma^2} + {\hat{p}_0}^2\,\sigma^2\right)\,j^2\,\sigma^4\,\cos^2\theta\right)}{d\,\hat{\tau}}.
\end{eqnarray}

This allows us to define the quantity
\begin{equation}
{\hat{C}_0}^2=\hat{L}_\theta + \frac{{\hat{L}_\phi}^2}{\sin^2\theta} - (k/\sigma^2 + p_0^2\sigma^2)\,j^2\,\sigma^4\,\cos^2\theta, 
\end{equation}
which is conserved along the motion. Furthermore, subtracting the conserved quantity $L^2_\phi$, we obtain a new conserved quantity
\begin{eqnarray}
\hat{C}^2&=&{\hat{C}_0}^2 - {L_\phi}^2, \\
&=&{\hat{L}_\theta}^2 + {\hat{L}_\phi}^2\,\left(\frac{1}{\sin^2\theta} -1\right) -k\,j^2\,\sigma^2\,\cos^2\theta \nonumber\\
&&- \hat{p}_0^2\sigma^6\,j^2\,\cos^2\theta\,,\\
&=&{\hat{L}_\theta}^2 + \left(\frac{{\hat{L}_\phi}^2}{\sin^2\theta} - \left(\frac{k}{\sigma^2} + \hat{p}_0^2\sigma^2\right)\,j^2\sigma^4\right)\cos^2\theta, \nonumber \\
&=&{\hat{L}_\theta}^2 + \cot^2\theta\,{\hat{L}_\phi}^2 - 2\,j^2\sigma^4\,\hat{E}_N\,\cos^2\theta,
\label{eq:Carter_pth}
\end{eqnarray}
which is the Carter constant. In the last step, we have used  $k=-1$, and expressed $\hat{p}_0$ in terms of $\hat{E}_N$, see Eq.~(\ref{eq:p0-EN}).

We can rewrite this constant in terms of the four velocity component $u^\theta$:
\begin{equation}
\hat{C}^2=\hat{r}^4\,{\Delta_{0+}}^2
{\hat{u}^\theta}{}^2  + \left(\frac{{\hat{L}_\phi}^2}{\sin^2\theta} - (k/\sigma^2 + \hat{p}_0^2\sigma^2)\,j^2\,\sigma^4\right)\,\cos^2\theta,  \label{eq:Carter_uth}
\end{equation}
and will be useful to express explicitely the $\hat{L}_\theta$ and the $\hat{u}^\theta$ components in terms of such constants:
\begin{eqnarray}
{\hat{L}_\theta}^2&=&\hat{C}^2 - \cot^2\theta\,{\hat{L}_\phi}^2 + 2\,j^2\sigma^4\,\hat{E}_N\,\cos^2\theta, \label{eq:Lth-CEN} \\ \hat{u}^\theta&=&\frac{\sqrt{\hat{C}^2 - \cot^2\theta\,{\hat{L}_\phi}^2 + 2\,j^2\sigma^4\,\hat{E}_N\,\cos^2\theta}}{\hat{r}^2\,\Delta_{0+}}.   \label{eq:uth-CEN}
\end{eqnarray}

Recall that $C=m\,q_0\,R_0\,\hat{C}$, as long as the Carter's constant has units of angular momentum.

\section{Testing the expression for the Black hole's mass determination} \label{sec:BH_Mass}

One of the foundations of the Newtonian Mechanics, is the derivation of the Kepler's laws from the Universal gravitational law. Kepler's third law is the simplest way to determine the mass of the central body based on observations of the apsides and the period of the orbit: 
\begin{equation}
M=\frac{\pi^2\,\left( \hat{r}_1+ \hat{r}_2\right)^3}{2\,G\,T^2}.  \label{eq:M_K3aL}
\end{equation}

In terms of the dimensionless quantities previously defined 
\begin{equation}
{\sigma^2}_{\rm 3rdKL}=\frac{\pi^2\left(\hat{r}_1 + \hat{r}_2 \right)^3}{2\,{\hat{T}}^2}. \label{eq:sigma_3aLK}
\end{equation}

We have presented two new expressions for determining $\sigma$, by means of the tangential velocity at the apsis: considering that we are in a Schwarzschild background, Eq.~(\ref{eq:sigma_Schw}) and considering that we are within the Keplerian description, Eq.~(\ref{eq:sigma_K}). 

Our code allows us to test the accuracy of these different expressions for the mass even if we are dealing with a Kerr background. Due to the precession on both angles, we define the period of a trajectory by the amount of time (time steps) it takes for the particle to start at a give apsis, $\hat{r}_2$ say, and returnt to the same value of $\hat{r}_2$ (although the angles will be, in general, different). Testing with a background of a black hole, with  Sagittarius A* like mass, and angular momentum $j=0.95$, we set the apsis at $\hat{r}_1=45, \hat{r}_2=1800$, and by changing the fiducial distance, we analyze equivalent orbits which are closer or farther from the black hole. That is, we set the fiducial distance as $n_{R0}\,R_{01UA}$, where $R_{01UA}$ stands for one atronomical unit, and consider several values for $n_{R0}$. In Tab.~\ref{tab:sigma}, we present our results for the percentage difference comparing the obtained value of $\sigma$ with the actual value of the input
\begin{table}  
\begin{center}
\begin{tabular}{|l || c | c | c |}  
\hline
 ${n_R}_0$ & $\Delta\,\sigma_{\rm S}\,\%$ &  $\Delta\,\sigma_{\rm K}\,\%$ & $\Delta\,\sigma_{\rm 3rdKL}\,\%$\\ 
\hline
0.1  & $0.11040656$  &  $-0.78581173$ & $0.12694847$ \\ \hline
1  & $0.00359792$  &  $-0.08519791$ & $0.01285743$ \\ \hline
10  & $0.0001129$  & $-0.00875656$  & $0.00086556$ \\ \hline
100  & $2.045193\,\times\,10^{-6}$  &  $-0.00088479$ & $0.00026655$ \\ \hline
\end{tabular}
\caption {Percentage difference between the computed $\sigma$ and the one from the input, $\sigma_{\rm S}$ stands for the case using the expression, Eq.~(\ref{eq:sigma_Schw}); $\sigma_{\rm K}$ for the case of our procedure within the Keplerian description, Eq.~(\ref{eq:sigma_K}), and $\sigma_{\rm 3rdKL}$ for the case when using Kepler's thir law, Eq.~(\ref{eq:sigma_3aLK}). We are considering equatorial orbits. The values ${n_{R}}_0$ are the multiples of the fiducial distance $R_0$ and has the effect of maintaining the shape of the trajectory but moves it closer or far from the black hole.}
\end{center} 
\end{table} \label{tab:sigma}

From Table~IV one can see that the three expressions give an adequate value for the mass of the central object, even though it is a Kerr background with $j=0.95$. The expression for the mass in the case where the apsis are at $\hat{r}_2=4.5\times\,10^3\,{\rm AU}$ and $\hat{r}_1=1.8\,\times\,10^6\,{\rm AU}$. 

\section{Testing of the code in the Solar System: Mercury and Earth} \label{sec:Mercury}

Even though the spacetime is considered to be described by the Kerr spacetime, the Newtonian results are correctly described with our code, a fact that allows us to test its performance and precision.

In these lines, we give the conditions so that the Solar System, and the Earth trajectory, be described within our code and we meassure the number uf time steps it takes to complete a revolution, and use the average of these data to compute the mass of the central object, {\it i. e.} the Sun, using the Kepler's third law, Eq.~(\ref{eq:M_K3aL}). In terms of our procedure, we write $M=n\,M_\odot$, 
obtaining:
\begin{equation}
n=\frac{\pi^2\,\left(\hat{r}_2+\hat{r}_1\right)^3}{2\,\sigma^2\,\hat{T}^2}=5.0082\,\times\,10^{8}\frac{\left(\hat{r}_2+\hat{r}_1\right)^3}{\hat{T}^2}, \label{eq:n-3K}    
\end{equation}
where we have used the value of the Solar mass, one astronomical unit for $R_0$ and $c, G$ in units mentioned in the text. We use our code giving the value of the apsis for the Earth, $\hat{r}_p=0.9859766005, \hat{r}_a=1.020067995$, in terms of Astronomical Units, {\it i. e.} ${n_R}_0=1$, and compute the average of the time steps needed to complete one turn.

\begin{table}
\begin{center}
\begin{tabular}{|l || c | c |}  
\hline
\multicolumn{1}{|c||}{Turn} & \multicolumn{1}{c|}{ Elapsed time $\hat{T}_{\rm Ap}$}& Interval\,$\times\,10^5$\\ 
{}& steps & steps \\
\hline
1   & $63585.00400609102$ & $63585.00400$\\ \hline
2   & $127170.00801218204$ &  $63585.00399$ \\ \hline
3   & $190755.01201827306$ &  $63585.0040$ \\ \hline
4   & $254340.01602436407$ &  $63585.0040$ \\ \hline
5   & $317925.0200304551$ &  $63585.0040$ \\ \hline
6   & $381509.38818650605$ &  $63584.3682$ \\ \hline
7   & $445094.39219259704$ &  $63585.0040$ \\ \hline
8   & $508679.3961986881$ &  $63585.0040$ \\ \hline
9   & $572264.4002047791$ &  $63585.0040$ \\ \hline
\end{tabular}
\caption {Time steps to complete a revolution. Apsis at $\hat{r}_p=0.9859766005, \hat{r}_a=1.020067995$, with ${n_R}_0=1$, so that the distances are multiples of one AU, and correspondingly, the fiducial time is $T_0=1.5802032141333376\,\times\,10^{-5}\,{\rm years}$. We are giving $10^8$ steps in our run.} \label{tab:T-Earth}
\end{center}
\end{table}

From the data presented in Table~(\ref{tab:T-Earth}), we obtain that the average time steps for a period is $\bar{\hat{T}}=63584.93333$, giving a time in years of $T=1.004771160\,{\rm years}$, and using
this value, together with ${n_R}_0=1$, $rp=0.98, ra=1.002$, in Eq.~(\ref{eq:n-3K}), we obtain a mass of $n=1.000000215$, (in this case $\sigma=9.926393360\times\,10^{-5}$, and we are considering that $j=0$), giving us a precision of our code of $-2.15\,\times\,10^{-5}\%$. 

Regarding the trajectory of Mercury, The turning points are $\hat{r_2}=0.30749\,\,{\rm AU}=46\,\times\,10^6\,{\rm km}$, and $\hat{r_1}=0.46669\,{\rm AU}=69.818\,\times\,10^6\,{\rm km}$, that is, in terms of the fiducial radius, $\hat{r}_1=7.78$, and $\hat{r}_2=11.82$ The Keplerian period, which in our units takes the expression, 
\begin{equation}
\hat{T}_K=2\,\pi\,\left(\frac{\hat{r}_1+\hat{r}_2}{2}\right)^\frac32, \label{eq:Kep3}
\end{equation}
equals in this case to $T_K=192.91$, in multiples of the characteristic time, $T_0$, that is, $T_K=87.74$ Terrestrial days, which is close to the known period for Mercury.
\begin{figure}
    \centering
    \includegraphics[scale=0.6]{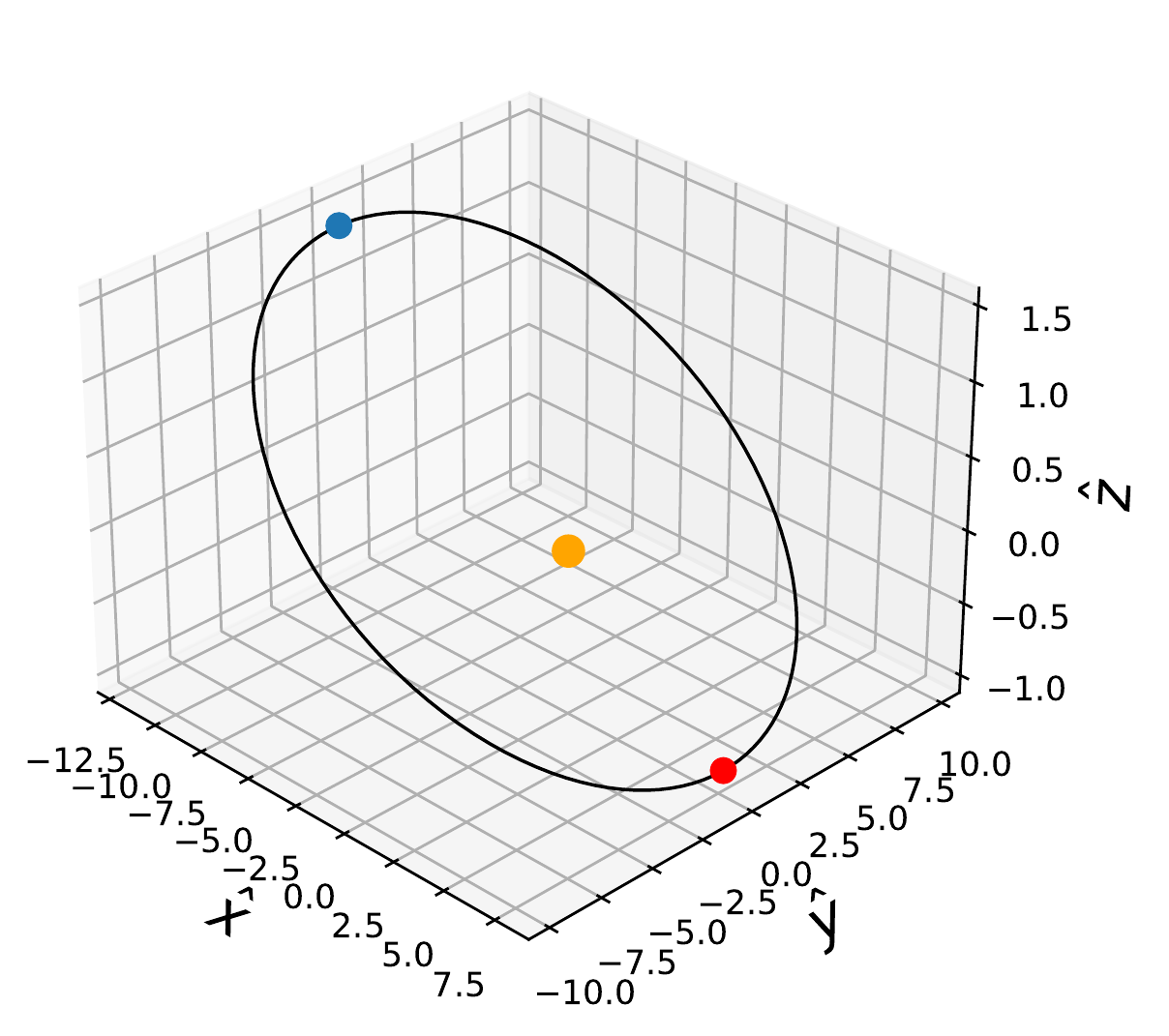}
    \caption{Plot of Mercury's trajectory, with $R_0 =5.906\,\times\,10^6\,{\rm km}$, so that the
    apsis are $\hat{r}_1=7.78$, and $\hat{r}_2=11.82$, $\theta_a=\frac{\pi}{2}-7\,\frac{\pi}{180}$, and the final time is $\hat{t}_f=386$. The bigger point denotes the position of the Sun, the smaller points mark the turning points.}
    \label{Apcides-N1}
\end{figure}

We can also compute the precession of the orbit as
\begin{equation}
\Delta\,\phi|_{\rm apsis}=3\,\pi\,\sigma^2\,\frac{\hat{r}_1+\hat{r}_2}{\hat{r}_1\,\hat{r}_2}, \label{eq:precM}    \end{equation}
and in this case, with $j=0$, as the expression was derived for a Schwarzschild spacetime, so that $\sigma=\sqrt{\frac{1}{2\,\cal N}}=5\,\times\,10^{-4}$ and the apsis for Mercury in units of $R_0$, we obtain that $\Delta\,\phi=5.02\,\times\,10^{-7}\,\frac{\rm rads}{\rm turn}$, which in 415 turns gives the known result of $42.98$ seconds of arc per century.  We can simulate such change in the angle $\phi$ at which the object reaches the apsis in each turn. This simulation was carried out considering that Mercury circled the sun 10 times with $10^8$ steps.

\acknowledgments

We are grateful to M. in Sc. Ismael Oviedo for enlightening discussion at the beginning of this project.  We also acknowledge Prof. C. Will suggestions on the applicability of our procedure to be used in programs regarding tests of the General Relativistic Theory. This work was partially supported by DGAPA-UNAM through grants IN110218 and IN105920, by the CONACyT Network Projects 
No. 376127 ``Sombras, lentes y ondas gravitatorias generadas por objetos compactos astrof\'\i sicos", and No. 304001``Estudio de campos escalares con aplicaciones en cosmolog\'ia y astrof\'isica". Also by the European Union’s Horizon 2020 research and innovation (RISE) program H2020-MSCA-RISE-2017 Grant No. FunFiCO-777740. 
LOV, ERC and VJ acknowledge support from the  CONACYT Graduate  Grant Program. CM acknowledges support from PROSNI-UDG.




\bibliographystyle{JHEP}
\bibliography{References.bib} 





\end{document}